# Phosphorus-rich stars with unusual abundances are challenging theoretical predictions

Thomas Masseron[1,2], D. A. García-Hernández[1,2], Raúl Santoveña[3,4], Arturo Manchado[1,2,5], Olga Zamora[1,2], Minia Manteiga[3,6], Carlos Dafonte[3,4]

**Abstract**

**Almost all chemical elements have been made by nucleosynthetic reactions in various kind of stars and have been accumulated along our cosmic history. Among those elements, the origin of phosphorus is of extreme interest because it is known to be essential for life such as we know on Earth. However, current models of (Galactic) chemical evolution under-predict the phosphorus we observe in our Solar System. Here we report the discovery of 15 phosphorus-rich stars with unusual overabundances of O, Mg, Si, Al, and Ce. Phosphorus-rich stars likely inherit their peculiar chemistry from another nearby stellar source but their intriguing chemical abundance pattern challenge the present stellar nucleosynthesis theoretical predictions. Specific effects such as rotation or advanced nucleosynthesis in convective-reactive regions in massive stars represent the most promising alternatives to explain the existence of phosphorus-rich stars. The phosphorus-rich stars progenitors may significantly contribute to the phosphorus present on Earth today.**

## Introduction

Life on Earth is mainly depending on five (chemical) elements: C, N, O, S and P. While P (phosphorus) plays a crucial role for deoxyribonucleic acid (DNA) and is the primary constituent in energy exchange in cells, in the form of adenosine triphosphate (ATP) and adenosine diphosphate (ADP)[1,2], phosphorus is the less abundant element of all five in the Solar System[21]. Yet the origin of phosphorus remains unclear. Since the publication of Burbidge, Burbidge, Fowler & Hoyle[4] in 1957 (aka B$^2$FH), we know that almost all chemical elements that are present in the Universe come from various generations of stars. Chemical evolution models trace the history of these elements all along the Galactic history by relying on a broad set of nucleosynthetic theoretical yields (mostly from stars) and by taking into account various stellar formation rates and initial stellar mass distributions. The resulting calculations are compared with observed elemental abundances derived from stellar spectra over various metallicities, Galacto-centric distances and stellar populations, including our Sun. While the most recent chemical evolution models successfully predict many elemental abundances across the Galaxy[5,6,7], they underestimate the (rare) observations of phosphorus (Fig. 1); in particular at low metallicity where all the available models systematically predict about three times less phosphorus than observed. This suggests that the main nucleosynthetic source of phosphorus is neglected in the models of chemical evolution or even possibly unknown.

Phosphorus is extremely difficult to measure in stellar spectra because its lines are inherently weak and are only available in the near-IR or UV spectral ranges, which very few high-resolution spectrographs can actually cover. The astronomical restrictions are such that only few recent studies[1-13] have succeeded in measuring phosphorus in other stars than the Sun. Fortunately, the APOGEE spectrograph[14] (see Methods: Searching for P-rich stars) operates in the near-IR (H-band) and thus offers a unique opportunity to detect and measure phosphorus from stellar spectra. Here we present the discovery of chemically peculiar stars, showing very high phosphorus abundances and for which there are currently no theoretical predictions that can explain such chemical pattern.

**Results**

*Detection of 15 Phosphorus-rich stars*

From the APOGEE spectral catalog containing now several hundreds of thousands of Galactic stars (mainly red giants), we could extract 15 objects that show clear P I spectral absorption features (Fig. 2 and Fig. 3). Once these P-detected stars have been identified, we run a full detailed chemical abundance analysis, deriving the main stellar parameters (i.e. $T_{eff}$, *log g*, [Fe/H], etc.; Supplementary Table 1) and P abundances together with the chemical abundances of many elements as possible also present in the APOGEE spectra. In Fig. 4, we show the resulting chemical abundance pattern along with the one observed in a control sample composed by twin stars (i.e. stars with very similar stellar parameters but with very weak phosphorus absorption features; the P-normal stars hereafter; Supplementary Table 2). It is very striking that the P-detected stars have not only very large P abundances but also enhanced O, Mg, Si, Al and Ce as compared to the P-normal stars that display the chemical abundance pattern typical of the well-known field stars (Fig. 4). The [P/Fe] abundance ratios are in the range 1.2 to 2.2 dex (i.e. ~10-100 larger than the Sun value on a logarithmic relative metallicity scale) with a typical estimated uncertainty of 0.1 dex (Supplementary Table 3). The radial velocities and the metallicity range (-1.05<[Fe/H]<-0.75) of P-rich stars suggest that these stars belong to the old stellar population of the Galactic thick disk or inner halo. Furthermore, the available photometric and distance data for the P-rich stars are consistent with low-mass giants (~1 $M_\odot$) with luminosities less than 1700 $L_\odot$ (Supplementary Table 1). Such low-mass stars are not expected to produce themselves P, O, Si, Mg, Al or Ce.

Regarding the statistics, our sample represents ~0.8% of the stars if we consider their number in the APOGEE data release relatively to all giants with similar temperatures and metallicities. We stress that this may represent a lower limit because of our detection methodology (see Methods: Searching for P-rich stars) and because this new stellar kind is potentially part of a larger family of Si- or Al-rich stars[15,16]. Whereas their relative chemical inhomogeneity clearly demonstrate that P-rich stars are not all born from the same gas cloud, one could wonder if these stars comes from a specific population. The orbital analysis links the P-rich stars with the Galactic inner halo (see Methods: Orbital analysis), thus excluding an extra-Galactic origin. A globular cluster (GC) origin can also be easily ruled out as GC stars systematically show Mg and O depletion along with Al and Si enhancement; conversely to the chemical abundance pattern of the P-rich stars.

*The possible progenitors of the P-rich stars*

So, where this peculiar chemical pattern comes from? In principle, there could be several possibilities to explain the extremely peculiar chemical pattern of P-rich stars: i) They could have accreted the material from an invisible companion. The most famous binary systems at such low metallicities are the so-called CH stars[17]; stars that have accreted large amounts of carbon from a now extinct asymptotic giant branch (AGB) star. But despite the metallicity and luminosity similarities with the CH stars, all our P-rich stars with observations spanning larger than a year

(with the exception of only one star) show radial velocity variations no higher than the instrument precision (500 m/s) ( Supplementary Table 1), advocating against the binary hypothesis; ii) They could be the result of a merging process with a companion where no radial velocity variations are expected. However, given that a merging event with advanced nucleosynthesis - like the one displayed by the P-rich stars - is likely to lead to a rather massive object, we argue that such a merged-star would be either much brighter or extinct by now; iii) They could be post-asymptotic giant branch[18] (post-AGB) stars. This possibility resides in the idea that part of the material that has been present in the stellar gas has condensed into dust such that some elements appear now depleted while the non-condensed elements seem enhanced in comparison. This phenomenon has not only been observed in post-AGBs but also in their descendants, the planetary nebulae[19], where dust is usually distributed in a circumstellar disk. No P-rich star, however, is as luminous as post-AGBs nor does the infrared photometry show any excess that would betray the presence of a circumstellar disk (Fig. 5; see also Methods: SEDs of P-rich stars). Also, Si, Al, Ni and Ca are refractory elements along with Fe and, in the gas-to-dust condensation scenario, they should all be depleted in a similar way; something that is clearly not observed in the P-rich stars. Therefore, the only left possibility is that P-rich stars were born from gas whose composition has been previously polluted by a progenitor with a quite specific and peculiar nucleosynthesis. However, we believe this peculiar source to be relatively nearby (i.e. a low dilution factor) to allow to supersede the already high initial chemical content of the interstellar medium at [Fe/H] ~-1.0.

According to chemical evolution models, there are two main stellar types that are responsible for phosphorus production in the Galaxy: low-mass (1-3 $M_\odot$) AGB stars[20] and massive stars (10-300 $M_\odot$)[21,22,23] that end their life as core-collapse supernovae. In the first type -the low-mass AGBs- $^{31}$P is produced after neutron-capture on $^{30}$Si. In such stars, large enhancements of C and heavy neutron-rich (s-process) elements like Ce are also theoretically predicted by the stellar nucleosynthesis models (Fig. 6; see also Methods: Comparison with nucleosynthesis predictions). Actually, such chemical pattern is well known to match and explain the existence of the above-mentioned CH-stars. Although phosphorus has never been measured in CH-stars, the chemical pattern of the P-rich stars does not resemble the CH-stars one; nor their known radial velocity variations. Thus, these arguments allow us to rule out the low-mass AGB progenitor. In contrast, the subsolar metallicity combined with the inner halo orbital motion indicate that P-rich stars were formed early in the Galaxy; so that only short time events (e.g. ~10-100 Myr) could have generated their peculiar chemical abundance pattern. Consequently, the most natural progenitor candidates would be massive stars. In this stellar type, $^{31}$P is, in principle, expected to be produced by a complex network of nuclear reactions during the oxygen-burning phase, although the so-called odd-even effect[6] systematically under-produce P compared to their neighboring elements Si and S (Fig. 6; see also Methods: Comparison with nucleosynthesis predictions). In an attempt to explain the solar phosphorus values, some previous studies have questioned and empirically modified the nuclear reaction rates assumed in the massive stars models[5], in order to artificially increase the P yields in the Galactic chemical evolution models. For example, the uncertainty on the $^{30}$P(p,γ)$^{31}$S reaction[24] may plague the final yields of phosphorus. Nevertheless, the observations of a supernova remnant[25] seem to confirm that P production in massive stars is consistent with the theoretical model yields, thus suggesting that the assumed nuclear reactions rates are reasonably accurate. Moreover, the fact that we observe at the same metallicity both P-rich and P-normal stars (Fig. 1) proves that inaccurate nuclear rates cannot be blamed alone as it would imply a systematic offset in the P yields but not a population split. Some alternative effects may then be considered to boost the production of P (and other odd elements) in massive stars.

The inclusion of rotation in the massive stars models[26,27] represents a very interesting effect. Indeed, the difference between models with and without rotation (Fig. 7) highlights some overproduction of P. The N enhancement in the P-rich stars would also be a signature of additional mixing due to the rotation effect, but it is not clear whether this enhancement is genuinely from the progenitor or if it is due to internal mixing of the P-rich giant star itself. The predicted production of P, however, is relatively modest (in comparison to the measured abundance in the P-rich stars) and – depending on the model – it does not always overcome the odd-even effect. In addition, although all massive stars models with rotation do not always provide similar yields, they systematically overproduce C, Na and S, which are not enriched in the P-rich stars (Fig. 4 and Fig. 7).

Another possible effect for the production of odd elements in massive stars involves advanced nucleosynthesis in convective-reactive regions and, in particular, in O-C shell mergers[28]. Although simulation of such nucleosynthesis requires expensive hydrodynamical computations, some models[28] include a prescription of such effect and provide P production factors for various masses (Fig. 7, bottom panel). Depending on the amount of C ingested in the O-shell, P can be abundantly produced under these conditions together with Al, Si, Mg and O; well in line with the P-rich star chemical pattern observed. The O-C shell mergers mechanism also boosts up the production of other odd elements like K and Sc[28]. However, no P-rich star shows any significant K enhancement (Fig. 4). Also, no Sc enhancement is detected in the only P-rich star (2M13535604+4437076; the brightest P-rich sample star) for which an optical spectrum is available (Fig. 8; see also Methods: Optical spectrum of a P-rich star). However, all models produce large amounts of C and Na as in models with rotation, which again contradicts the observations.

*The heavy element abundance pattern of a P-rich star*

Actually, further analysis of the 2M13535604+4437076 optical spectrum provides additional hints, especially when considering the neutron-capture elements. In Fig. 8, we display the full chemical abundance pattern (both from the optical and the H-band spectra) of the P-rich star 2M13535604+4437076, which is overplotted with the solar-scaled r-process[52] pattern and theoretically predicted s-process patterns[27,53] for a comparison with the heavy neutron-capture elements observed in P-rich stars. Assuming that this star is representative of all the other P-rich stars, the enhancement of the first-peak elements (namely Rb, Sr, Y, Zr) and the second-peak ones (Ba, La, Ce, Nd) together with the low Eu value seem to indicate that the nucleosynthesis in the P-rich stars is neither compatible with high-neutron density processes such as the solar-scaled r-process nor it is compatible with the weak s-process general pattern[27,29]. The absence of Cu and Zn enhancement further disfavors the occurrence of the weak s-process nucleosynthesis. In contrast, the values of [Rb/Fe] and [Sr/Fe] are very similar, usually indicating a relatively high neutron density. But even more intriguing is the really high overabundance of Ba compared to any other neutron-capture element (see Methods: Optical spectrum of a P-rich star). The s-process neither the r-process are able to predict such a large Ba enhancement ([Ba/La] = +0.7) compared to the other second-peak elements (La or Ce), and this observation (if confirmed as an intrinsic property of these peculiar stars) could help to distinguish the P-rich stars progenitors (or a still unknown nucleosynthesis process). In any case, while the lack of a clear r-process pattern could rule out core-collapse supernovae[30], other effects such as rotation may play a role and alleviate this apparent heavy element inconsistency. Although the predictions of the massive stars models do not agree with each other (Fig. 7), some models that include rotation[26] allow strong overproduction of the heaviest elements including Ce in the pre-supernovae stages by the combination of increased neutrons and low metallicity. Nevertheless, the same authors warned about the variable impact of explosive nucleosynthesis on the pre-explosion yields. Explosive nucleosynthesis is expected to favor the production of elements beyond Si and up to the heavy elements first peak[30]. During the explosion, Si production and the r-process nucleosynthesis are triggered; the latter favoring the production of more neutron-capture elements. However, the P-rich stars do seem to not show any Ca, Ti or Fe-peak elements (Co, Ni, Cu or Zn) enhancement that is also expected by explosive nucleosynthesis. Therefore, if the P-rich stars progenitor was a massive star, the explosion yields have been reduced such as it is for the most massive models with strong fallback (e.g. the $25 M_\odot$ model of ref.[29] in Fig. 7).

Another hypothesis to account for the production of the heavy elements beyond the Fe-peak, considers the nucleosynthesis in compact objects, and in particular neutron stars mergers[31]. In theory, the dynamical ejecta in such objects are extremely rich in neutrons, thus favoring the production of the heaviest nuclei; i.e. with A>130 (Ba, La, Ce, Nd) as it is seen in the P-rich stars. Similarly, models of magnetorotationally driven supernovae[32] are also able to produce large amounts of second and third peak r-process elements. However, the low Eu abundance - a typical heavy r-process element - measured in the P-rich star 2M13535604+4437076, cast serious doubts on the possible occurrence of a neutron star merger event or a magnetorotationally driven supernova.

More exotic P producers such as super-AGB stars and Novae could be invoked to explain the existence of P-rich stars. However, they are actually discarded because of clear mismatches between the model predictions and the abundances of P-rich stars, among other caveats (see Methods: Exotic nucleosynthesis).

In summary, although there are some promising scenarios, none of the current theoretical predictions for standard stellar nucleosynthesis can consistently reproduce all the chemical abundances in P-rich stars. Hence, the discovery of P-rich stars demands a broader exploration of stellar nucleosynthesis networks within specific conditions such as rotation or convective-reactive events.

**Discussion**

Still, until the true nature of the P-rich stars polluters is revealed, we can speculate that they could have appreciably contributed to the Galactic chemical evolution (and de facto to the Solar System and Earth).

The existence of chemically abnormal stars such as the P-rich stars at metallicities ~-1.0 provides interesting hints in the Galactic chemical evolution framework. Chemical inhomogeneities in heavy elements have been observed at metallicities lower than [Fe/H]<-2.0. This fact has been used[33-36] to infer the progenitor/s of the heavy elements and to argue (via chemical evolution modelling) that neutron star mergers, magnetorotationally driven supernovae or rotating massive stars would be adequate candidates. Given the lack of any binary evidence, the existence of P-rich stars can be considered as another inhomogeneity signature (Fig. 9), although of a distinct type from the one cited above given their relatively high metallicity ([Fe/H]~-1.0). Interestingly, inhomogeneities at such high metallicities have never been observed until now, suggesting that either the mixing scale of the P-rich progenitors is not large in order to overcome the already enriched interstellar medium or that we only observe the most enhanced P-rich stars, the remaining being too diluted to be chemically distinguishable. In addition, to explain that only a certain fraction of stars at [Fe/H]~-1 are P-rich, a specific distribution of progenitor types could also be invoked. Indeed, the mean Galactic chemical evolution of several elements – including neutron-capture elements – has been satisfactorily reproduced by adopting an empirical distribution of rotation in massive stars[7]. Similarly, the inclusion of a crude constant rate of only 10% O-C shell mergers in a simple Galactic chemical evolution model has permitted to reproduce the solar P abundance[28]. This demonstrates that even a minority population such as the P-rich stars population could appreciably contribute to the global chemical enrichment. It is to be noted here that, along with a too low P solar abundance, some chemical evolution models[7] coincidentally underpredict Mg, Al and some Si isotopes in the Early Solar System.

By combining the arguments above on the chemical evolution and the chemical patterns by stellar nucleosynthesis, we could tentatively suggest that a combination of specific effects (i.e. rotation and/or advanced nucleosynthesis in convective-reactive regions) in massive stars may represent a promising way to explain our discovery of the chemically peculiar P-rich stars. However, before validating such scenario, there still remain strong contradictions between the nucleosynthesis models available and the chemical abundance pattern observed in P-rich stars. In our view, the recurrent discrepancy between the observations and all the models reside in finding a source or nucleosynthesis processes that allow strong P production but also O, Mg, Al, Si and Ce but with none or negligible C and Na enhancement.

## Methods

*Searching for P-rich stars*

The spectra used in the present work were obtained from the SDSS-IV/APOGEE survey DR14[37]. The APOGEE survey is a large spectroscopic survey taking high-resolution (R~22,500) spectra of F-, G-, and K-type stars in the H-band using the Apache Point Observatory telescope. All the data have been reduced in a homogeneous manner[38] and publicly available. In principle, the data are also analyzed via an automatic tool[39], thus providing effective temperatures ($T_{\text{eff}}$), surface gravities ($\log g$) and elemental abundances for each stellar spectrum. However, given that this pipeline is based on a pre-computed synthetic grid of model spectra, elemental abundances beyond the grid ranges such as observed in the P-rich stars are not valid. Therefore, we have developed a new algorithm to search for specific enhanced absorption/emission features in the APOGEE spectra database. This code has been previously validated against the Ce lines of the APOGEE spectra. The algorithm extracts the peaks in the wavelengths of interest (here the strongest and less blended P I lines at 15711.5 and 16482.9 Å.), categorizing and filtering each detection by means of different thresholds defined based on the local values of the signal. For each P line of interest, a window for local analysis is created. First, the code looks for peaks of absorption (or emission) close to the wavelength of each P line, establishing a threshold calculated from the signal continuum and the variance, so that the algorithm is sensitive to spectra with noise or very ripple. Secondly, the area of the peak is calculated, with the aim of filtering the P lines with little width and/or height. The detection of a P line is considered positive when both steps are met. A visual inspection of the candidates (~50) was finally made, ending with a final sample of 15 P-detected stars. It is to be noted here that our detection method is rather restrictive and there are probably much more P-enhanced stars to be found. For the same reason we cannot presently know whether the relatively narrow metallicity range of the P-rich stars is real or as a result of a selection bias of our detection algorithm.

*Chemical abundance analysis*

Once the P-detected stars were identified, the effective temperatures were derived using a J-K-$T_{eff}$ relation[40], and surface gravities (log$g$) were determined by isochrones[41]. With the $T_{eff}$ and log$g$ values fixed, we derive (when possible) the abundances of C, N, O, Na, Mg, Al, Si, P, S, K, Ca, Ti V, Cr, Mn, Co, Ni and Ce as well as the metallicity, macroturbulence, and v sini for each P-detected star. For this, we used the Brussels Automatic Code for Characterizing High accUracy Spectra (BACCHUS)[42] with the atomic/molecular APOGEE linelists[43] as input. In short, BACCHUS basically uses MARCS model atmospheres[44] and the spectral synthesis code Turbospectrum[45] to find the best model (synthetic) spectra on a line-by-line basis. The oscillator strengths (log$gf$) for the two main P I lines at 1.57115 and 1.64829 μm are -0.404 and -0.273, respectively. They were checked to reproduce satisfactorily the P I spectral lines in the Sun with an assumed solar abundance of 5.36[1]. The illustration of the best fits around the P lines in the P-rich stars is shown in Fig. 2 and Fig. 3. We note that the 1.64829 μm line is blended with a CO molecular line. Fortunately, other CO lines are present nearby at ~1.6493 μm, whose good fit guaranties that the blend is properly taken into account and that the P abundance is secure. Moreover, we could also derive P abundance in a few of the best spectra of P-normal stars (e.g. Fig. 1) which lead to a value of [P/Fe]~+0.6 in agreement with literature studies[13]. We also must emphasize that the N abundances are measured via the CN molecular lines and thus are strongly dependent on the O and C abundances because of molecular equilibrium. Consequently, the N abundances displayed in Supplementary Tables 1 and 2 with only upper limits on O may be overestimated. The systematic errors on abundances were derived by assuming the effective temperatures and surface gravities as provided by the APOGEE automatic pipeline[39]. The good agreement between the photometric and spectroscopic temperatures as shown in Supplementary Tables 3 and 4 suggests that our parameters determination is fairly robust. Random errors on abundances were evaluated by the line-by-line abundance dispersion. We note that nearly all P-rich stars APOGEE spectra have high signal-to-noise (S/N) ratios (> 70) and the majority of stars more than 100 (see Supplementary Table 1). In parallel, a set of control (twin) stars with similar $T_{eff}$, log$g$ and metallicities as the P-rich stars and with high S/N ratios (> 150) were selected from the APOGEE DR14 catalog[38] (the so-called P-normal stars). We rigorously applied the same methodology for the stellar parameters and chemical abundances determination to the P-normal (twin) stars. The stellar parameters and abundances of the P-rich and P-normal stars and all related errors are shown in Supplementary Tables 1 and 2, respectively. In Fig. 9, we compare the abundances obtained for both the P-rich and P-normal samples against literature[46]. The chemical abundance pattern of the P-normal stars is similar to field (Galactic thick disk and halo) stars, as expected. The agreement between the P-normal stars and literature is overall good although some slight systematics may appear for Si, Ni or Al. While those systematics can easily be attributed to various assumptions in both our analysis and in the literature (e.g. temperature scales, linelists or 3D and NLTE effects), they remain small in comparison to the difference between the P-normal stars and the P-rich ones, and thus not affecting our conclusions.

*Optical spectrum of a P-rich star*

In parallel to the APOGEE spectrum, we have obtained high-resolution (R~67,000) optical spectra for the brightest P-rich star in our sample (2M13535604+4437076). The spectra were obtained with the FIES spectrograph[47] at the 2.5m Nordic Optical Telescope (La Palma, Spain) in January and February 2020 under the Spanish service time. Four single spectra were acquired with a total exposure time of 4 hours, covering the 4200-9000 Å spectral range. The data reduction followed the standard FIES pipeline and the S/N achieved in the final summed spectrum is 30 at 4500 Å and 60 at 9000 Å. The chemical abundance analysis was done consistently with the APOGEE H-band spectral analysis; i.e. we rigorously adopted the same stellar parameters, model atmosphere and program codes. The atomic/molecular linelists used for the optical part are those used in the Gaia-ESO survey[48]. The final abundances are shown in Fig. 8 together with the IR-derived abundances. Although the abundances from the optical spectra suffers from larger uncertainties due to the lower S/N ratio, there is a good agreement between the abundances obtained from the H-band spectrum and the optical one, with the exception of Al and K. The latter discrepancies can be explained by the strong non-local thermodynamic equilibrium (NLTE) effects that, respectively, affect the Al lines in the H-band[38] and the K lines in the optical[49]. Thus, the Al optical abundance is certainly the most accurate, while for K it is the near-infrared one. We also note that for O, two values where available from the O I forbidden transition line at 6300 Å and from the O triplet at ~7777 Å. The latter indicator shows ~0.1 dex less oxygen than the former one, which we assume is due to NLTE effects. Therefore, we adopted the O abundance from the 6300Å transition.

The high overabundance of Ba compared to any other neutron-capture element deserves some attention here. The high Ba content that we find is not likely a saturation effect, as this would tend to underestimate the Ba abundance rather than overestimate it. However, NLTE effects are known to affect the Ba abundances for the optical lines we have been using (4554, 5853, 6141, and 6496 Å). We can estimate that theoretical NLTE corrections[50,51] could be only up to -0.3 for the stellar parameters of the P-rich star 2M13535604+4437076. Moreover, our measurement method consists in looking only at the line wings and not the equivalent widths; the line core is much more sensitive to both saturation and NLTE. Thus, our Ba measurement is quite NLTE free and we conclude that the large Ba overabundance such that [Ba/La] = +0.7 is quite robust. In any case, optical follow-up observations of more P-rich stars would be useful to prove/disprove if the high Ba is an intrinsic property of the P-rich stars.

Finally, we have attempted to measure Li in this star but only an upper limit of $\log\varepsilon(Li)=1.2$ dex could be obtained. Unfortunately, this cannot represent a constraining limit regarding the initial Li of the progenitor because the observed star, according to its stellar parameters, is very likely a red clump star, where any pristine Li has been altered in situ and even possibly fully destroyed.

*SEDs of P-rich stars*

We have collected the photometry - available via the Virtual Observatory Sed Analyzer (VOSA) online tool[54] - for each P-rich star and draw their respective spectral energy distribution (SED) in Fig. 5. Reddening corrections were applied to all filters assuming reddening values provided by Bayler-Jones et al. from Bayesian inference on Gaia data[55]. The synthetic spectra displayed in this figure have been tailored with the final stellar parameters and abundances using the same model atmosphere code MARCS[44] and the radiative transfer code Turbospectrum[45] as for the chemical abundance analysis (see above). The luminosities reported in Supplementary Tables 1 and 2 were derived by combining distances (from the Gaia mission[56] for distances less than 2 kpc and from Bayesian inference[55]) and bolometric corrections[57]. The Gaia parallaxes were corrected by - 0.029 mas as recommended by the Gaia team[58]. Concerning the radial velocities shown in Supplementary Tables 1 and 2, the values were simply taken from the published APOGEE DR14 catalog[38].

*Orbital analysis*

Orbital properties of the sample stars have been extracted from the the astroNN catalog of abundances and distances for APOGEE DR16 stars. Properties of the orbits in the Milky Way such as eccentricities, peri/apocenter radii, maximal disk height z, orbital actions, frequencies and angles (and their uncertainties) for all stars are computed using the fast method[59] assuming the MWPotential2014 gravitational potential[60]. The range of values for the P-rich stars is

$Z_{max}$=[3,15]kpc and $e$=[0.6,1.0], all corresponding to regular inner halo orbits[15,61]. Moreover, it has been established that in the metallicity range [-1.5,-0.5] one can distinguish any stream (including the Gaia-Enceladus stream) from the in situ stars based on chemistry, in particular alpha elements such as Mg and Ca. While Mg is not reliable in the P-rich stars because it has been clearly spoiled by an unidentified progenitor, the Ca seems to indicate that the P-rich stars belong to the original Galactic population and has not been accreted from an external population.

*Comparison with nucleosynthesis predictions*

In Fig. 6, we compare the chemical abundance pattern observed in P-rich stars with the nucleosynthetic theoretical predictions (collected from up to date literature) for low- and intermediate-mass AGBs[20], supernovae[22], pair-instability supernovae[62], type Ia supernovae[63], super-AGBs[64] and novae[65]. The published model yields were not computed with the exact same initial composition (metals) as it was at the epoch of the formation of the P-rich stars and usually assume a simple solar scaled pattern. Consequently, in order to mimic the real initial abundances of P-rich stars we added up the observed abundance pattern of the P-normal (twin) stars to the theoretical yields. We also stress that we use the pure theoretical yields in the comparison and we have neglected here any dilution (e.g. as it would be expected in a binary/merger system) that could have occurred since the formation of the P-rich star. However, this dilution factor would only tend to shrink down all the chemical abundance pattern together and it would thus not affect our discussion, mainly based on the relative elemental abundances.

For each progenitor type, we also display several models that reflect the variety of parameters that can influence the final yields. In particular, we show in the case of AGBs and supernovae (SNe) how the initial mass affects the final abundances. In the case of super-AGBs, we show also that the metallicity has a strong impact. For novae, on the other hand, we show that the white dwarf type (CO or ONe) has a strong effect. Naturally, other model prescriptions can influence the final theoretical yields such as the mass loss treatment (e.g. AGBs/super-AGBs), rotation or the energy of the explosion in the case of SNe. The variations in the model prescriptions lead to an even broader spectrum of nucleosynthesis models and stellar evolution codes in the literature. However, a detailed review of all of them is out of the scope of this paper. Instead, we base our main argumentation on the fundamental nuclear reactions nearly independent of the model prescriptions within each stellar type (e.g. the Na-O correlation for AGBs, the α-element production and odd-even effects in massive stars or the C and N production in super-AGBs and novae) (see main text and Methods: Exotic nucleosynthesis).

Nevertheless, we must note here that all the stellar abundances are bound to the quality of the nuclear reaction rates network used in the nucleosynthesis models. In parallel, it is very important that the analysis of the observations are equally accurate. Although we acknowledge that 3D and NLTE effects[66] need to be accounted for in order to certify that the analysis of the spectra are accurate, our differential approach against P-normal stars demonstrates the peculiarity of the P-rich stars pattern.

*Exotic nucleosynthesis*

There are two types of rare objects that can be found in literature and that are also able to produce a large amount of P: super-AGB[64,67] stars and novae[68]. Super-AGB (~6-10 $M_\odot$) stars theoretical models predict large amounts of P, O, Mg, Si and Al, although strongly depending on the model parameters (Fig. 6). Interestingly, this time $^{31}$P is enhanced by α-particle captures on $^{27}$Al but the largest amounts only occur at very low metallicities ([Fe/H]~-2.3); i.e. much lower than in P-rich stars ([Fe/H] ~ -1.0). No super-AGB stars have yet been unambiguously identified though, which may still leave room for some neglected/underestimated effect in the actual models predictions.

The enhanced O together with a lack of Na observed in the P-rich stars is, however, a quite crippling argument against a contribution from any AGB star, super or intermediate-mass type. Novae represent interesting other alternatives where their explosion events lead to intense H-burning reactions and where P is produced this time by capture of protons on Si. Very large amounts of P, O, Mg, Si and Al are predicted by the existing models, although the exact yields (again) strongly depend on the white dwarf type and on the thermonuclear conditions[24,65] (Fig. 6). The idea that novae could significantly pollute their surrounding environment and the chemical composition of the subsequent generations of stars has been recently revived in order to explain the content of $^7$Li in stars of the Galactic disk[69]. Furthermore, it has long been suspected that $^{15}$N in the Solar System has also originated in novae[70]. Actually, measurements of CNO isotopic ratios in pre-solar grains[71,72] have represented the best clue to testify that novae products have reached our Solar System. We could constrain $^{12}$C/$^{13}$C, $^{14}$N/$^{15}$N and $^{16}$O/$^{17}$O in the coolest P-rich star (respectively 3, >2 ,>5 ), but it is theoretically expected that they had brought to the surface their own C, N and O products (during the so-called first dredge-up) and had also probably undergone extra-mixing processes along the red giant branch. Such mixing processes imply that the original C, N and O isotopic ratios have been altered towards the standard values of a 1 M$_\odot$ metal-poor giant[73] and, consequently, any pristine C, N or O isotopic signature of a previous nova has been diluted. However, the main issue regarding the nova scenario resides in our observation of enhanced Ce in the P-rich stars. All novae models always consider that explosive H-burning nucleosynthesis do not affect the elements heavier than $^{40}$Ca[74], and none of the heavy elements such as Ce is expected to be enhanced. Other caveat to the Nova scenario is that the first nova events should occur quite late in the history of the Galaxy. This is because a white dwarf needs to be formed beforehand. In contrast, the P-rich stars display quite low metallicities, suggesting that they are rather old stars. Finally, the lack of large enhancements of C in the P-rich stars also contradicts all present novae and super-AGBs models.

Finally, although they are not expected to produce P, (Galactic) chemical evolution models usually take into account the theoretical yields of intermediate-mass (3-6 M$_\odot$) AGB stars[20,67] and thermonuclear supernovae (aka type Ia supernovae). However, an analysis of their typical nucleosynthetic pattern allow us to confirm that they are most probably not related to the P-rich stars' origin (Fig. 6). Indeed, intermediate-mass AGBs are characterized by the Mg-Al-Si nuclear reaction chain, which destroys Mg. Similarly, no type Ia supernovae are theoretically expected to produce large amounts of Fe-peak elements such as Fe, Mn or Cr, but no O or Mg, which abounds in our P-rich stars.

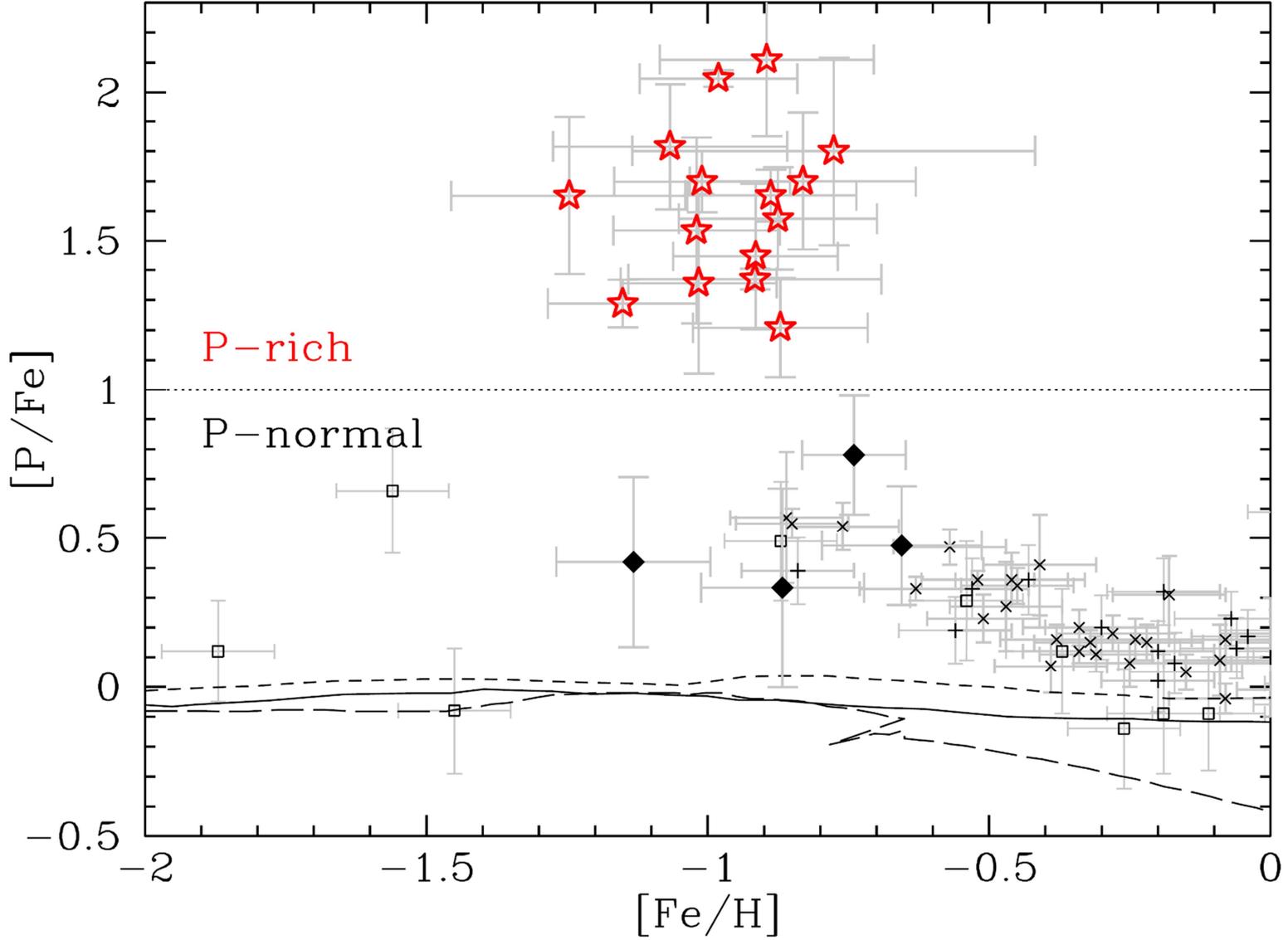

**Fig. 1. Phosphorus abundances as a function of metallicity.** Red stars and black diamonds show, respectively, the P-rich and P-normal stars analyzed in this work. Error bars represent measurement uncertainties such as shown in Supplementary Tables 2 and 4. The P-normal stars are the group of comparison (twins) stars with very similar stellar parameters to the P-rich ones but with a chemical abundance pattern similar to field stars (see Methods: chemical abundance analysis). The dotted line shows the empirical definition adopted here to distinguish the P-rich stars from the P-normal ones. Open squares, crosses, and plus signs are literature values from ref.[8], ref.[13] and refs.[11,12], respectively. The long-dashed line shows the chemical evolution model from ref.[5], short-dashed line from ref.[6] and solid line from ref.[7] Error bars have been adopted from same literature references.

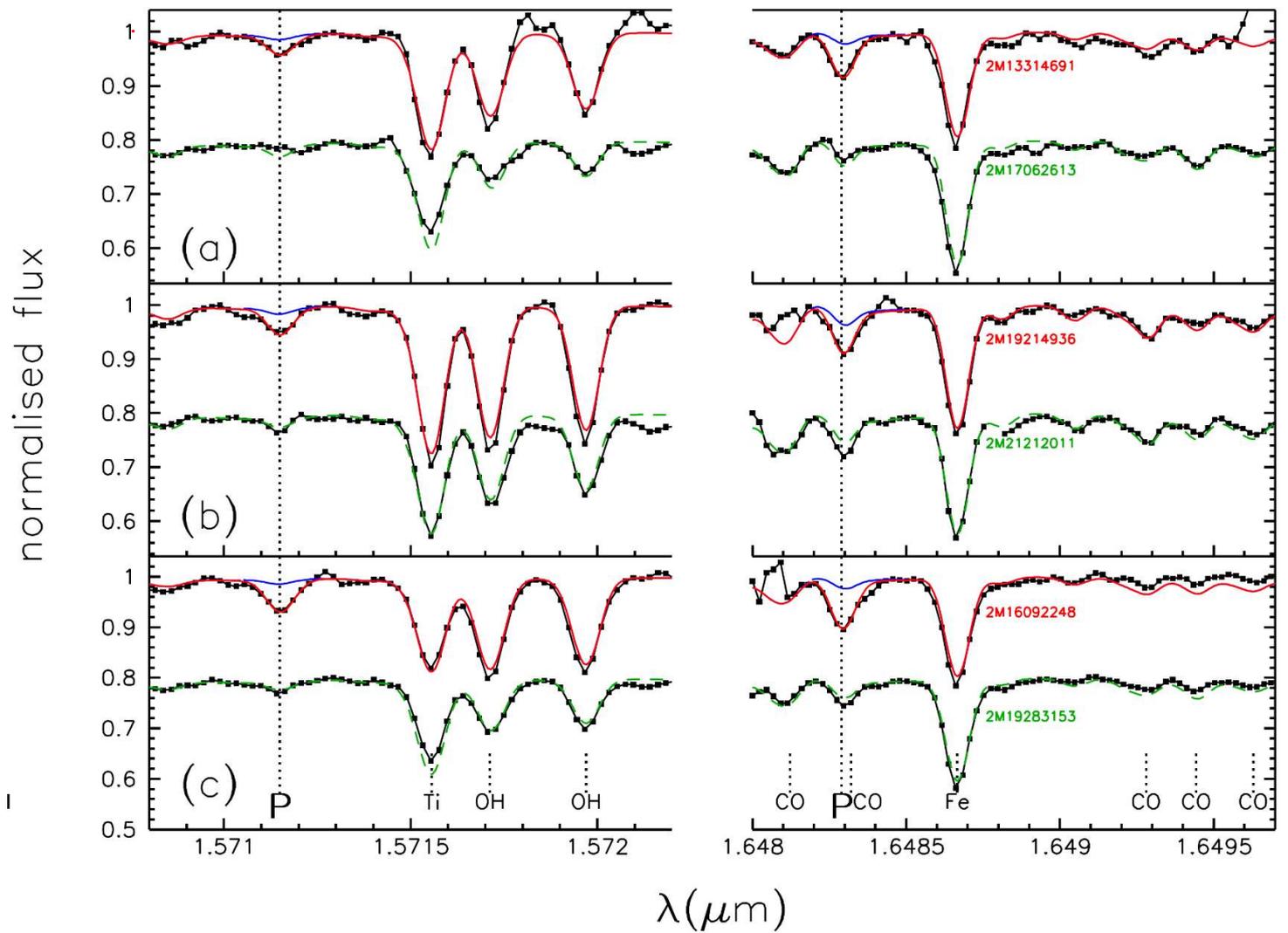

**Fig.2. Sample spectra of P-rich stars.** Observed H-band spectra (in black) of three P-rich stars and P-normal stars (panels (a), (b) and (c)) around the two P I absorption lines at 15711.5 and 16482.9 Å. A gap in between wavelengths has been introduced to highlight the wavelength discontinuity. The best model spectra (respectively in red for the P-rich stars and green for the P-normal stars) are shown together with model spectra without the P lines (in blue). The P-normal stars spectra have been offset by -0.2 in flux for display purposes.

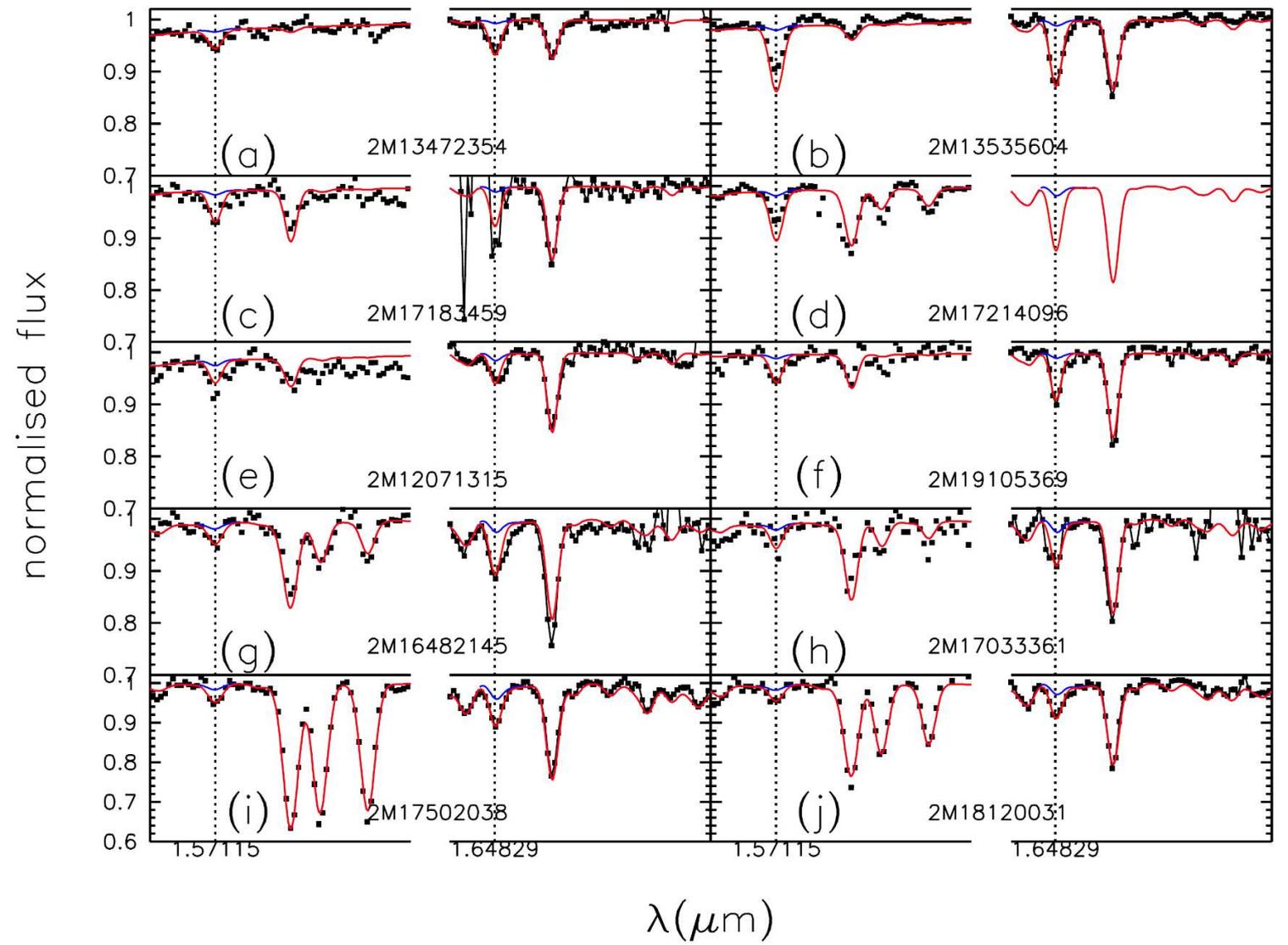

**Fig. 3. Spectra of additional P-rich stars in our sample.** Observed H-band spectra (in black) of additional (10, panels (a) to (j)) P-rich stars in our sample around the two P I absorption lines at 1.57115 and 1.64829 μm. A gap in between wavelengths has been introduced to highlight the wavelength discontinuity. The best model spectra (in red) are shown together with model spectra without the P lines (in blue).

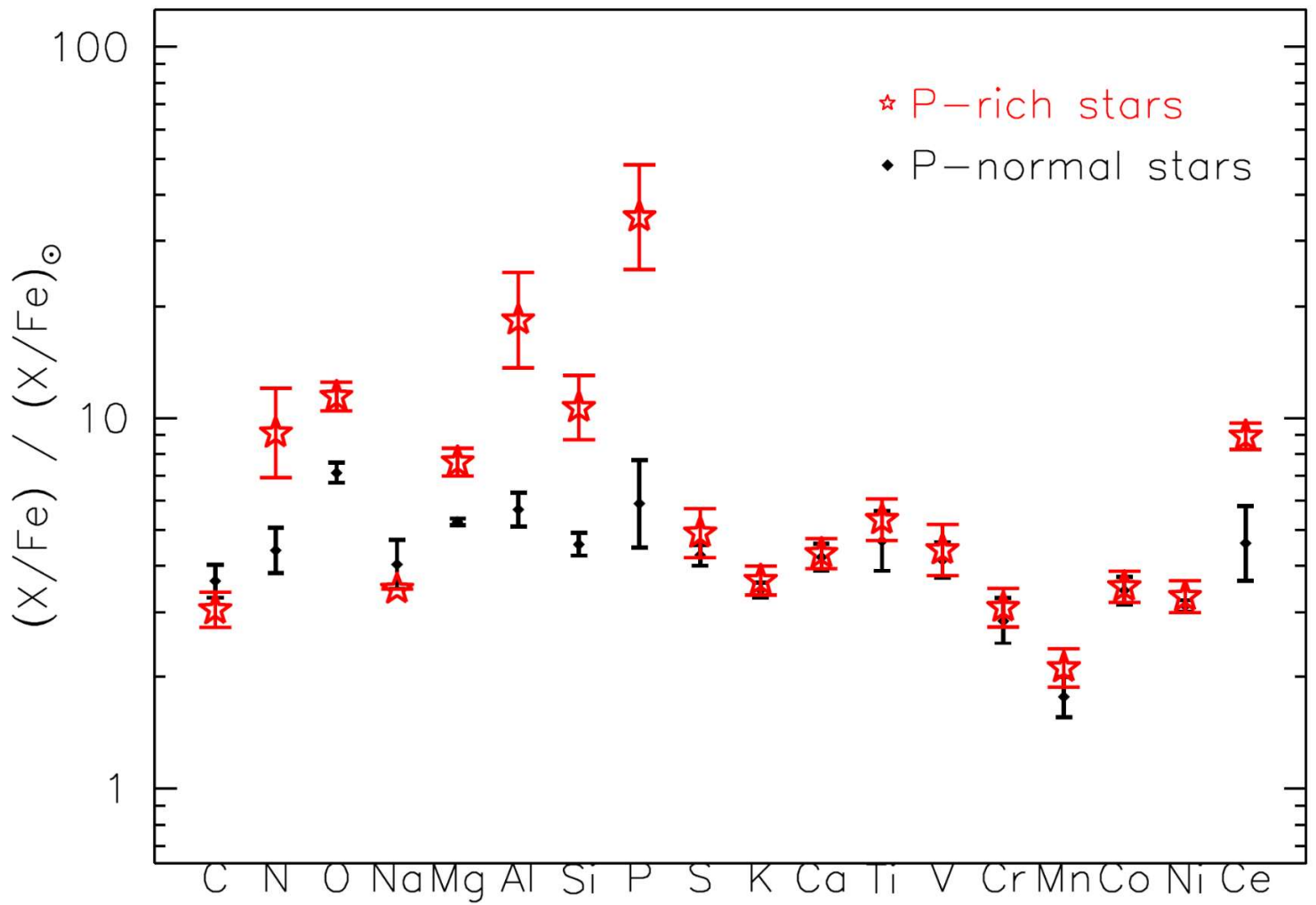

**Fig. 4. Chemical abundance pattern observed in P-rich stars.** Red and black symbols correspond to the median chemical abundances of the P-rich and P-normal (twin) stars, respectively. Error bars show the star-to-star abundance rms scatter. Note that only stars without upper limits are considered in the computation of the median values. The P-rich stars show clear enhancements of O, Mg, Al, Si, P and Ce compared to their normal (twin) counterparts.

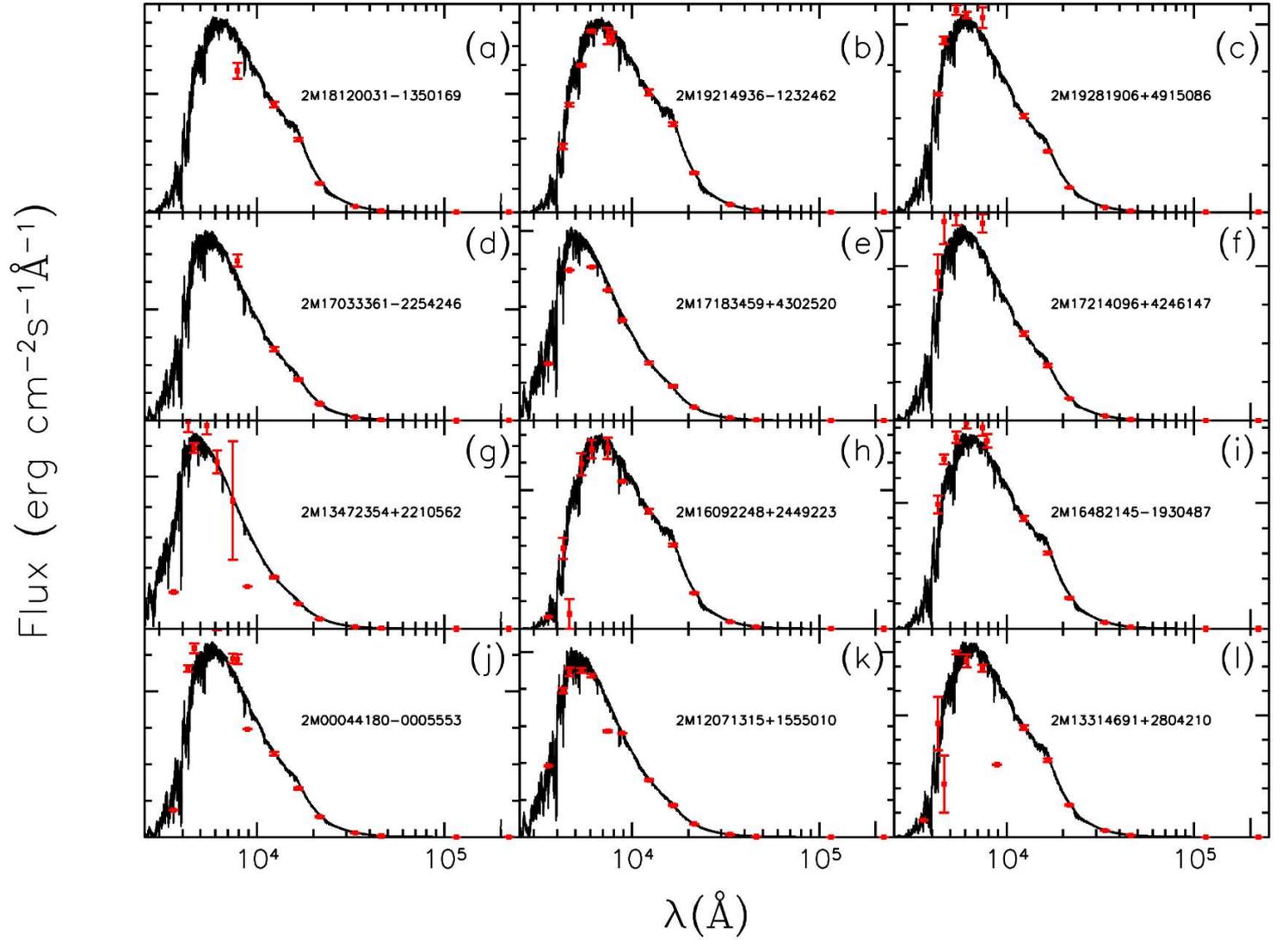

**Fig. 5**. **Spectral energy distributions (SEDs) of the P-rich stars**. The SEDs (broad band photometry from 2500 to 250000Å in red) for 12 P-rich sample (panels (a)-(l)) stars are shown in comparison with the synthesized fluxes (in black) with the derived stellar parameters and chemical abundances of each star. Error bars show the photometric measurement errors such as provided by VOSA.

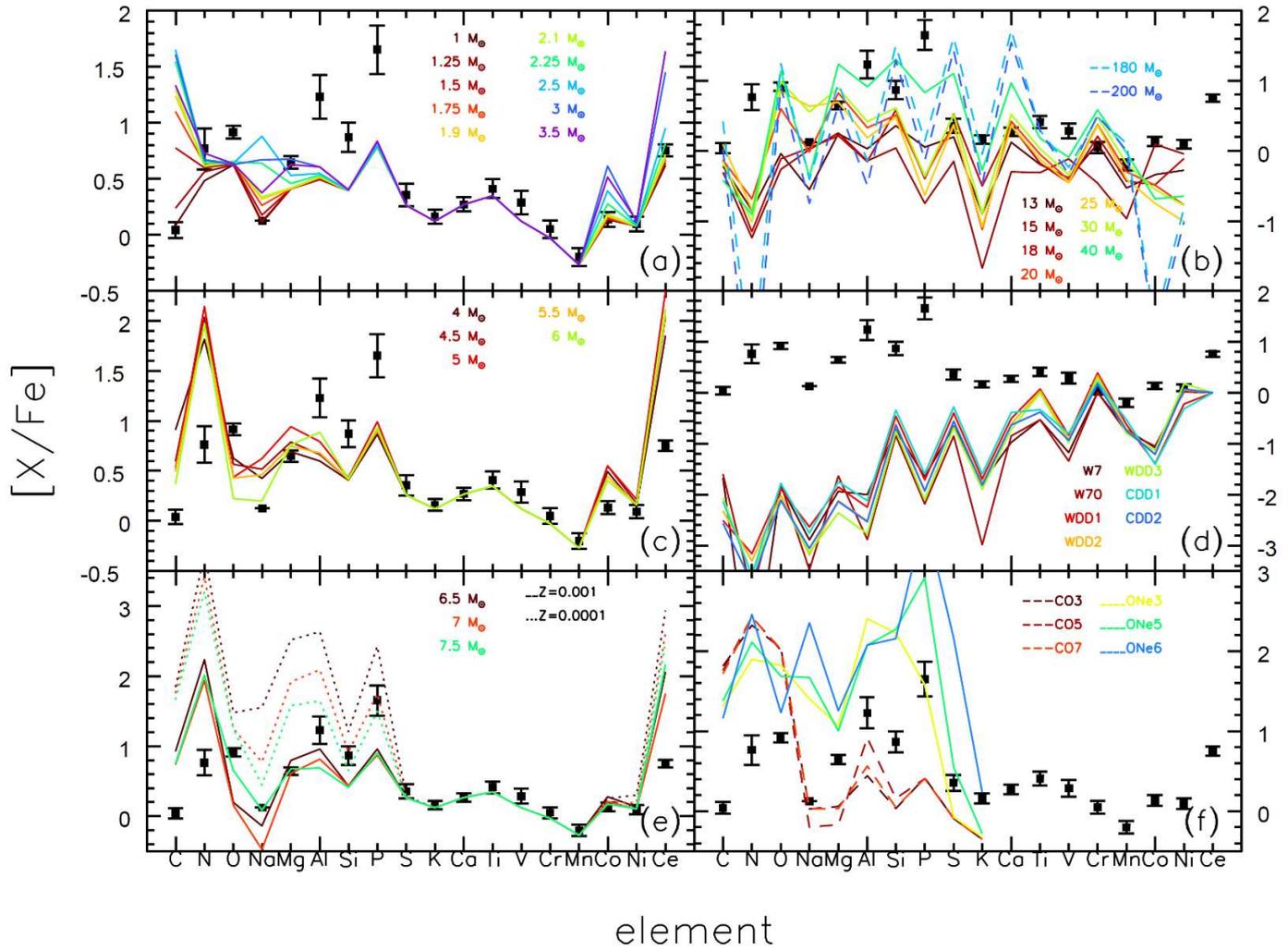

**Fig. 6. P-rich stars versus nucleosynthesis predictions.** Median chemical abundance pattern of the P-rich stars (black squares) against the several model prediction patterns (colored lines), where [X/Fe]=log10($n$(X)/$n$(Fe))-log10($n$(X)/$n$(Fe))$_\odot$. Error bars show the star-to-star abundance rms scatter. (a) The low-mass AGB subpanel shows the yields[20] for a metallicity of $Z$=0.004 (Fe/H]~-0.7) and various initial masses [1.0-3.0]M$_\odot$ in a rainbow fashion (steps of 0.5 M$_\odot$), the redder being the lower masses. (b) In the core collapse supernova (SNII) subpanel, we show standard models (i.e. without any specific effect like rotation or O-C mergers) where the mass range is [13-40]M$_\odot$ and metallicity such that $Z$=0.001 ([Fe/H] ~-1.3)[22]. In the same subpanel, the pair-instability supernovae (PISN) yields[62] are represented by dashed lines for masses of 180 and 200 M$_\odot$. (c)The initial masses of the intermediate-mass AGB predictions[20] (intM-AGB) range from 3.5 to 6 M$_\odot$ with $Z$=0.004 (or [Fe/H]~-0.7). (d) The SN Type Ia (SNIa) yields[63] cover all values in central densities (1.37x10$^9$-2.12x10$^9$ g cm$^{-3}$) and deflagration speeds (1.5%-5% of sound speed) provided by the authors. (e) Theoretical predictions for super-AGB stars[64] (S-AGB) at two metallicities ($Z$=0.001, 0.0001 or [Fe/H] ~-1.3, -2.3; continuous and dotted lines, respectively) and three initial masses (6.5, 7.5 and 8.0 M$_\odot$) are displayed. (f) Finally, we display the only solar metallicity ($Z$=0.014 or [Fe/H] ~0.0) novae yields available in the literature[65] with CO core WDs (dashed lines) and ONe WDs (continuous lines) with the same mass range of [0.85-1.15]M$_\odot$.

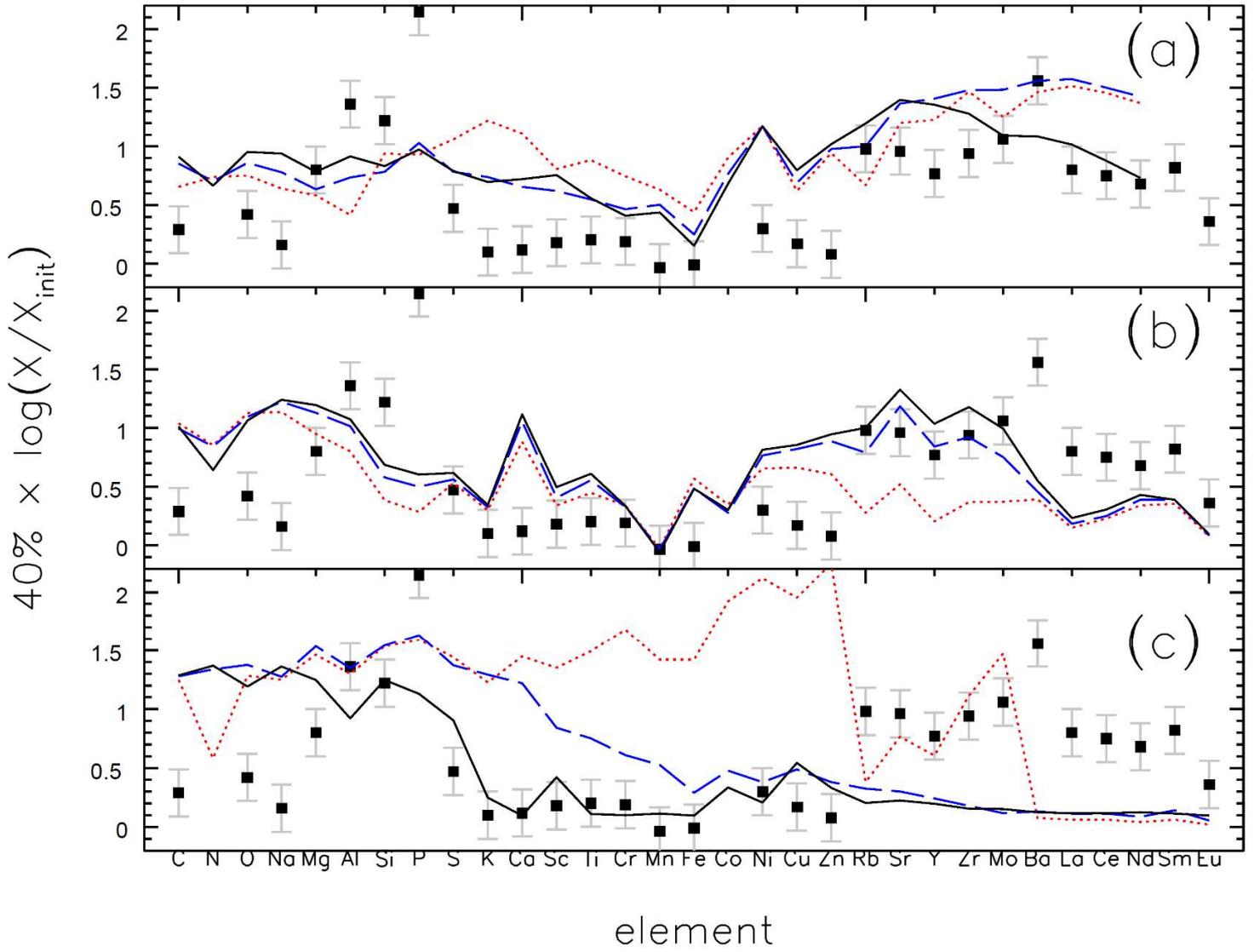

**Fig. 7. Predictions of overabundance factors for massive star models against the abundance pattern of the P-rich star 2M13535604+4437076.** Error bars show measurement uncertainties such as displayed in Supplementary Table 3. (a) $Z=0.001$, 15, 20, 25 $M_\odot$ models (respectively dotted red, long-dashed blue and solid black lines) from ref.[26] with rotation. (b) $Z=0.001$, 15, 20, 25 $M_\odot$ models (respectively dotted red, long-dashed blue and solid black lines) from ref.[27] with rotation. (c) $Z=0.001$ 15, 20, 25 $M_\odot$ models (respectively dotted red, long dashed blue and solid black lines) from ref.[28] with O-C shell mergers. A rough dilution factor of 40% has been applied to all model yields.

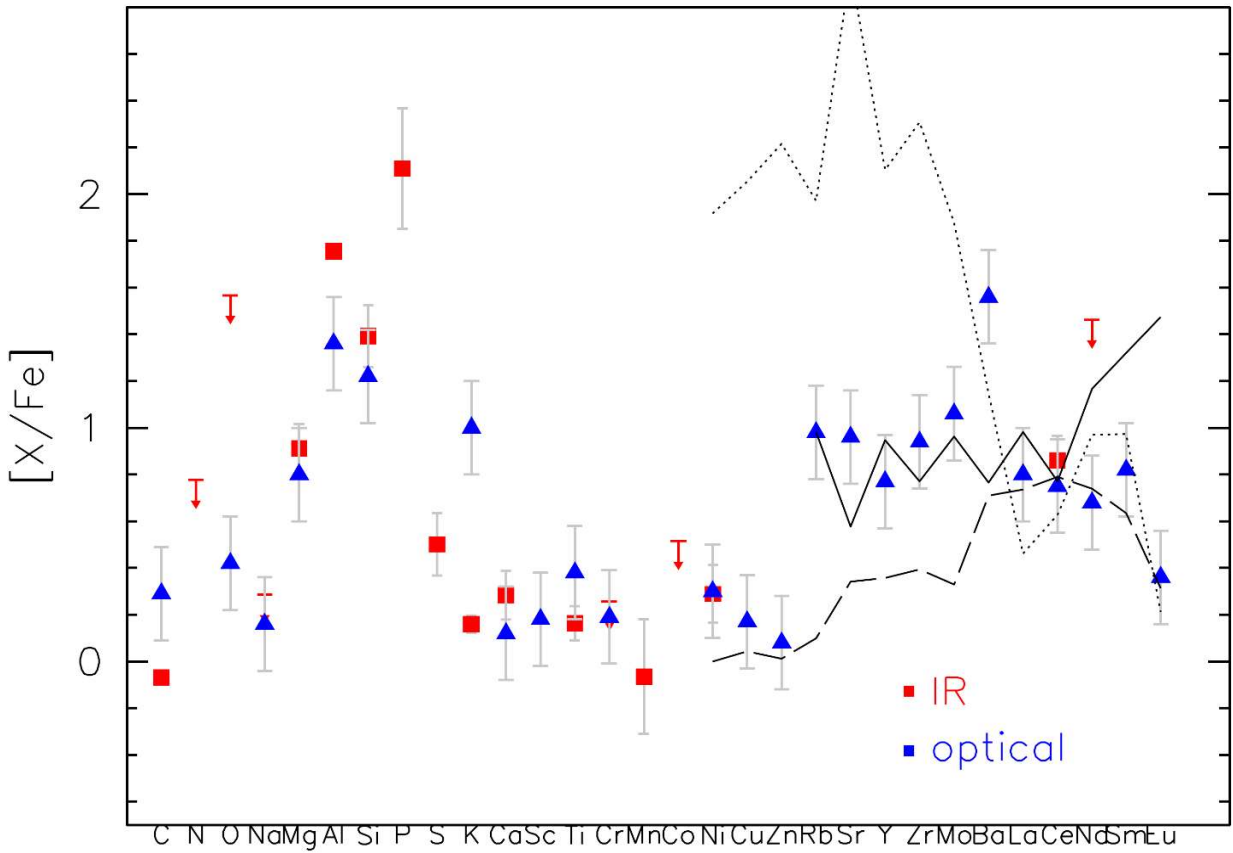

**Fig. 8**. **Chemical abundance pattern of the P-rich star 2M13535604+4437076.** Blue triangles and red squares are the abundances values obtained from the optical and near-IR spectra, respectively, where the arrows indicate upper limits. Error bars show the measurement uncertainties such as shown in Supplementary Table 3. The solid black line shows the solar r-process pattern scaled to match the Ce abundance from ref.[52], while the long-dashed line shows an s-process theoretical pattern for a 1.5M$_\odot$ star at [Fe/H] = -1.0 diluted by 50% to match the Ce abundance from ref.[53] (a model as close as possible to the P-rich star chemical pattern, as found in literature) and the dotted line illustrate the weak s-process such as obtained by a model of a 20 M$_\odot$ star with $Z$=0.001 without dilution from ref.[27].

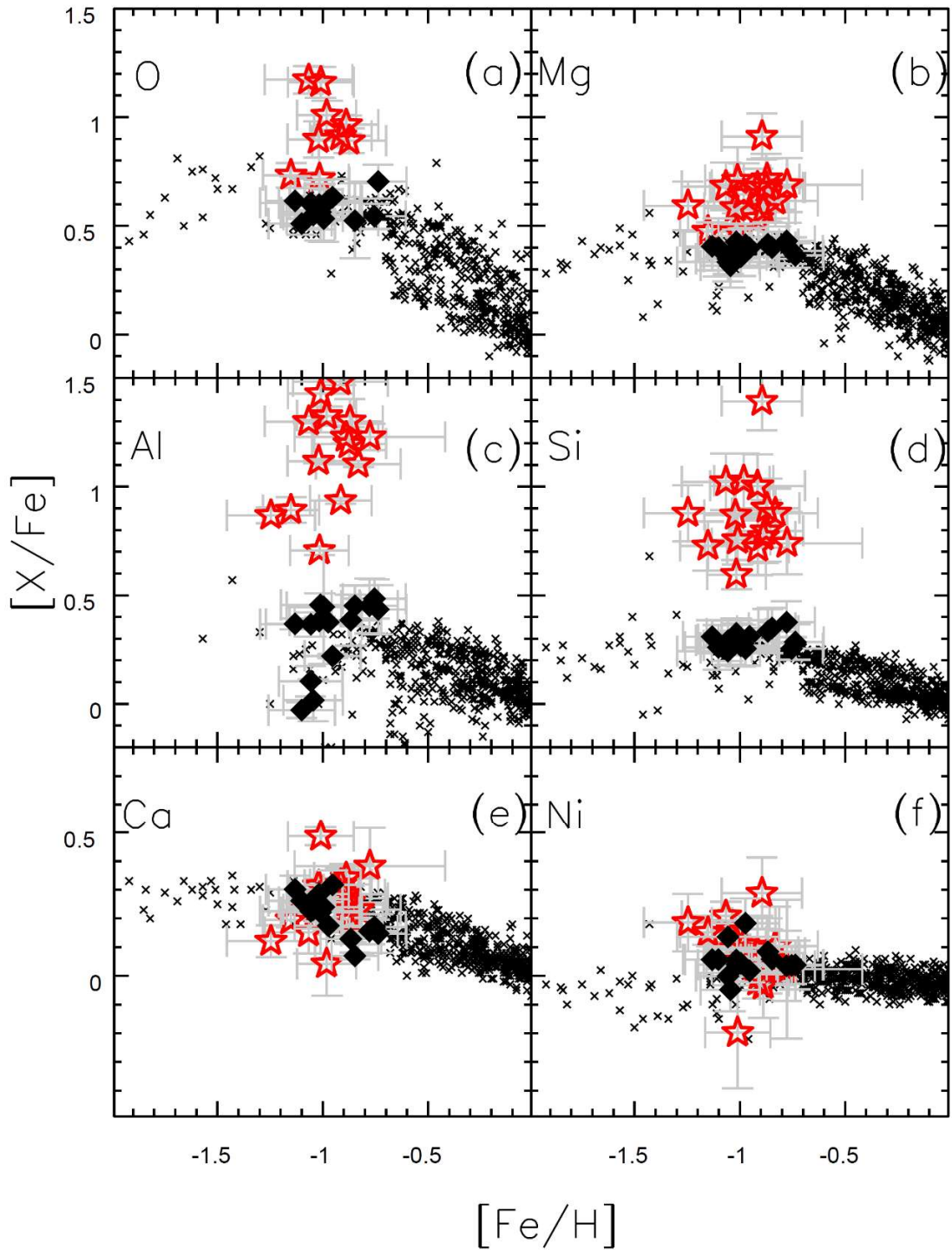

**Fig. 9. Elemental abundances as a function of metallicity.** (a) oxygen (b) magnesium (c) aluminium (d) silicon (e) calcium (f) nickel. The red stars and black diamonds show the P-rich and P-normal stars, respectively, while the black crosses correspond to the optical literature values for field dwarf stars[46]. Error bars indicate our measurement uncertainties such as displayed in Supplementary Table 3 and 4.

## Acknowledgments

We acknowledge support from the State Research Agency (AEI) of the Spanish Ministry of Science, Innovation and Universities (MCIU) and the European Regional Development Fund (FEDER) under grants AYA2017-88254-P and RTI2018-095076-B-C22. We also wish to acknowledge the support received from the Centro de Investigación de Galicia CITIC, funded by Xunta de Galicia and the European Union (European Regional Development Fund - FEDER Galicia 2014-2020 Program) by grant ED431G 2019/01, the research group fund ED431B 2018/42 and the scholarship ED481A-2019/155. This article is based on observations made in the Observatorios de Canarias del IAC with the Nordic Optical Telescope (NOT) operated on the island of La Palma by NOTSA in the Observatorio de Los Muchachos (ORM). This publication makes use of VOSA, developed under the Spanish Virtual Observatory project supported by the Spanish MINECO through grant AYA2017-84089. This work has made use of data from the European Space Agency (ESA) mission Gaia , processed by the Gaia Data Processing and Analysis Consortium (DPAC, ). Funding for the DPAC has been provided by national institutions, in particular the institutions participating in the Gaia Multilateral Agreement. Funding for the Sloan Digital Sky Survey IV has been provided by the Alfred P. Sloan Foundation, the U.S. Department of Energy Office of Science, and the Participating Institutions. SDSS acknowledges support and resources from the Center for High-Performance Computing at the University of Utah. SDSS is managed by the Astrophysical Research Consortium for the Participating Institutions of the SDSS Collaboration including the Brazilian Participation Group, the Carnegie Institution for Science, Carnegie Mellon University, the Chilean Participation Group, the French Participation Group, Harvard-Smithsonian Center for Astrophysics, Instituto de Astrofísica de Canarias, The Johns Hopkins University, Kavli Institute for the Physics and Mathematics of the Universe (IPMU) / University of Tokyo, the Korean Participation Group, Lawrence Berkeley National Laboratory, Leibniz Institut für Astrophysik Potsdam (AIP), Max-Planck-Institut für Astronomie (MPIA Heidelberg), Max-Planck-Institut für Astrophysik (MPA Garching), Max-Planck-Institut für Extraterrestrische Physik (MPE), National Astronomical Observatories of China, New Mexico State University, New York University, University of Notre Dame, Observatório Nacional / MCTI, The Ohio State University, Pennsylvania State University, Shanghai Astronomical Observatory, United Kingdom Participation Group, Universidad Nacional Autónoma de México, University of Arizona, University of Colorado Boulder, University of Oxford, University of Portsmouth, University of Utah, University of Virginia, University of Washington, University of Wisconsin, Vanderbilt University, and Yale University.

## Author Information

Affiliation:

[1] Instituto de Astrofísica de Canarias, E-38205 La Laguna, Tenerife, Spain

[2] Departamento de Astrofísica, Universidad de La Laguna, E-38206 La Laguna, Tenerife, Spain

T. Masseron (tmasseron@iac.es), D.A. García-Hernández (agarcia@iac.es), A. Manchado (amt@iac.es), O. Zamora (ozamora@iac.es)

[3] CITIC, Centre for Information and Communications Technology Research, Universidade da Coruña, Campus de Elviña sn, 15071 A Coruña, Spain

R. Santoveña Gómez (raul.santovena@udc.es), M. Manteiga (minia.manteiga@udc.es), C. Dafonte (dafonte@udc.es)

[4] Universidade da Coruña (UDC), Dept. of Computer Science and IT,
University of A Coruna, 15071, A Coruña

[5] Consejo Superior de Investigaciones Científicas, Madrid, Spain

[6] Universidade da Coruña (UDC), Department of Nautical Sciences and Marine Engineering, 15011 A Coruña, Spain

## Corresponding author

Correspondence to T. Masseron

# Phosphorus-rich stars with unusual abundances are challenging theoretical predictions

Thomas Masseron, D. A. García-Hernández, Raúl Santoveña, Arturo Manchado, Olga Zamora, Minia Manteiga, Carlos Dafonte

## Supplementary Tables

Supplementary Table 1: Basic parameters of the P-rich stars

| star | $T_{\rm eff}$ | $\log g$ | $\mu_t$ | [Fe/H] | SNR | dist. | Lum. | <rv> | rms | #visit | $\Delta t$ |
|---|---|---|---|---|---|---|---|---|---|---|---|
| 2M00044180-0005553 | 4597( 33) | 2.01(0.13) | 1.38 | $-1.01^{0.13}_{0.00}$ | 553 | 4300 | 276 | -59.67 | 0.08 | 10 | 300 |
| 2M12071315+1555010 | 5119( 29) | 2.24(0.28) | 1.33 | $-0.87^{0.15}_{0.01}$ | 55 | 9700 | 264 | 28.49 | 0.05 | 3 | 411 |
| 2M13314691+2804210 | 4224(177) | 1.22(0.52) | 1.66 | $-1.02^{0.14}_{0.03}$ | 249 | 9000 | 1120 | -1.80 | 0.03 | 4 | 8 |
| 2M13472354+2210562 | 5377(295) | 2.2(1.56) | 1.34 | $-1.24^{0.21}_{0.17}$ | 164 | 2200 | 60 | 30.75 | 0.47 | 3 | 3 |
| 2M13535604+4437076 | 5143( 97) | 2.5(0.82) | 1.30 | $-0.89^{0.19}_{0.02}$ | 346 | 1500 | 60 | -128.88 | 0.01 | 3 | 346 |
| 2M16092248+2449223 | 4208( 75) | 1.07(0.1) | 1.74 | $-1.06^{0.2}_{0.04}$ | 363 | 9800 | 1561 | -92.42 | 0.29 | 5 | 15 |
| 2M16482145-1930487 | 4481( 57) | 1.61(1.31) | 1.50 | $-0.87^{0.17}_{0.03}$ | 98 | 8000 | 330 | 65.93 | 0.04 | 3 | 706 |
| 2M17033361-2254246 | 4783( 84) | 2.27(0.03) | 1.33 | $-0.77^{0.35}_{0.02}$ | 62 | 8700 | 49 | 81.50 | 1.76* | 15 | 388 |
| 2M17183459+4302520 | 5130( 1) | 2.66(0.25) | 1.29 | $-0.91^{0.22}_{0.00}$ | 91 | 10900 | 92 | -135.80 | 0.27 | 15 | 371 |
| 2M17214096+4246147 | 4550( 67) | 1.49(0.06) | 1.54 | $-0.98^{0.14}_{0.02}$ | 179 | 9800 | 883 | -297.56 | 0.07 | 5 | 31 |
| 2M17502038-2805411 | 3885( 46) | 0.28(0.54) | 2.25 | $-1.01^{0.15}_{0.12}$ | 101 | 4300 | 249 | -52.73 | 0.13 | 2 | 11 |
| 2M18120031-1350169 | 4244(179) | 1.36(0.1) | 1.60 | $-0.88^{0.15}_{0.07}$ | 130 | 5000 | 200 | -36.24 | 0.00 | 1 | 0 |
| 2M19105369+2717150 | 5255( 31) | 3.21(0.14) | 1.32 | $-0.83^{0.2}_{0.03}$ | 73 | 4700 | 43 | -5.08 | 0.18 | 7 | 11 |
| 2M19214936-1232462 | 4096( 97) | 1.09(0.46) | 1.73 | $-0.91^{0.14}_{0.04}$ | 266 | 12300 | 1277 | -146.57 | 0.23 | 3 | 222 |
| 2M19281906+4915086 | 4425(109) | 1.63(0.15) | 1.49 | $-1.15^{0.13}_{0.05}$ | 374 | 4700 | 603 | -302.63 | 0.21 | 2 | 50 |

Effective temperatures ($T_{\rm eff}$ in K) and surface gravities ($\log g$) uncertainties are shown in parentheses. $\mu_t$ are the microturbulence velocities (in km s$^{-1}$) assumed for the abundances derivation. [Fe/H] represent the metallicities of the stars relative to the solar value based on Fe lines measurement. The upper indices show the random errors while the bottom indices show the systematic errors. dist are the distances of the stars in parsec. Lum. are the stars luminosities relative to the Solar one. $<$rv$>$ show the average radial velocities, rms the dispersion around that value (both in km s$^{-1}$) and #visit the number of observed spectra. and $\Delta t$ (in days) the total observation period range.
* star suspected to be a binary

Supplementary Table 2: Basic parameters of the P-normal stars

| star | $T_{eff}$ | log g | $\mu_t$ | [Fe/H] | SNR | dist. | Lum. | <rv> | rms | #visit |
|---|---|---|---|---|---|---|---|---|---|---|
| 2M00022364+1558539 | 5229( 79) | 2.4(0.88) | 1.31 | $-0.77^{0.13}_{0.01}$ | 301 | 1900 | 57 | -60.30 | 0.05 | 3 |
| 2M01262518+1719099 | 5232(130) | 2.39(1.03) | 1.31 | $-0.84^{0.11}_{0.01}$ | 724 | 1000 | 55 | 35.53 | 0.21 | 5 |
| 2M01271266+1801037 | 4730(216) | 2.17(0.21) | 1.35 | $-1.13^{0.13}_{0.1}$ | 1405 | 700 | 96 | 101.02 | 0.53 | 5 |
| 2M02150319-0437411 | 5027(177) | 2.9(0.08) | 1.29 | $-0.97^{0.16}_{0.1}$ | 550 | 700 | 20 | 80.77 | 0.10 | 3 |
| 2M07404920+3621206 | 4610( 54) | 2.26(0.28) | 1.33 | $-0.65^{0.13}_{0.00}$ | 358 | 2900 | 71 | 66.74 | 0.04 | 7 |
| 2M07461742+4828201 | 4709(182) | 2.03(0.25) | 1.37 | $-1.05^{0.24}_{0.11}$ | 410 | 2400 | 194 | -39.46 | 0.07 | 2 |
| 2M11202609+0024314 | 4391( 94) | 1.02(0.47) | 1.76 | $-1.1^{0.15}_{0.05}$ | 271 | 9800 | 846 | 195.36 | 0.30 | 3 |
| 2M13240275+2516183 | 5168(138) | 2.29(0.99) | 1.33 | $-0.86^{0.14}_{0.04}$ | 305 | 1300 | 69 | 69.02 | 0.04 | 3 |
| 2M15563661+2716483 | 4533(165) | 1.78(0.1) | 1.44 | $-1.01^{0.18}_{0.02}$ | 327 | 2700 | 315 | 17.36 | 0.01 | 2 |
| 2M16164586+4652131 | 4668(113) | 1.97(0.07) | 1.39 | $-1.01^{0.14}_{0.06}$ | 305 | 2500 | 140 | -112.77 | 0.04 | 3 |
| 2M16471103-0156177 | 4566(640) | 1.74(1.26) | 1.45 | $-0.99^{0.17}_{0.29}$ | 264 | 8700 | 2441 | -44.43 | 0.00 | 1 |
| 2M17142525+4324562 | 4490(110) | 1.7(0) | 1.47 | $-1.04^{0.14}_{0.05}$ | 406 | 5700 | 341 | -106.93 | 0.04 | 5 |
| 2M19311218-0840354 | 4765( 27) | 2.09(0.02) | 1.36 | $-0.65^{0.14}_{0.01}$ | 212 | 5200 | 211 | 18.51 | 0.28 | 3 |
| 2M19320870+4926497 | 4733(123) | 2.4(0.21) | 1.31 | $-0.74^{0.09}_{0.06}$ | 212 | 1600 | 54 | -80.09 | 0.00 | 1 |
| 2M21330531+0041464 | 4548( 42) | 2.05(0.19) | 1.37 | $-0.82^{0.13}_{0.03}$ | 476 | 1900 | 120 | -96.13 | 0.12 | 6 |

Effective temperatures ($T_{eff}$ in K) and surface gravities ($\log g$) uncertainties are shown in parentheses. $\mu_t$ are the microturbulence velocities (in km s$^{-1}$) assumed for the abundances derivation. [Fe/H] represent the metallicities of the stars relative to the solar value based on Fe lines measurement. The upper indices show the random errors while the bottom indices show the systematic errors. dist are the distances of the stars in parsec. Lum. are the stars luminosities relative to the Solar one. $<rv>$ show the average radial velocities, rms the dispersion around that value (both in km s$^{-1}$) and #visit the number of observed spectra.

Supplementary Table 3: Abundances of the P-rich stars

| star | [C/Fe] | [N/Fe] | [O/Fe] | [Na/Fe] | [Mg/Fe] | [Al/Fe] | [Si/Fe] | [P/Fe] | [S/Fe] |
|---|---|---|---|---|---|---|---|---|---|
| 2M00044180 | $0.24^{0.05}_{0.08}$ | $0.35^{0.14}_{0.09}$ | $0.71^{0.03}_{0.07}$ | $<0.42$ | $0.47^{0.1}_{0.04}$ | $0.7^{0.02}_{0.09}$ | $0.59^{0.06}_{0.02}$ | $1.35^{0.3}_{0.09}$ | $0.32^{0.04}_{0.08}$ |
| 2M12071315 | $0.06^{0.06}_{0.39}$ | $<0.91$ | $<1.25$ | $<0.27$ | $0.71^{0.09}_{0.09}$ | $1.3^{0.1}_{0.09}$ | $0.9^{0.13}_{0.04}$ | $1.2^{0.16}_{0.1}$ | $0.42^{0.14}_{0.11}$ |
| 2M13314691 | $-0.06^{0.17}_{0.16}$ | $0.77^{0.05}_{0.19}$ | $0.89^{0.08}_{0.29}$ | $<0.31$ | $0.58^{0.09}_{0.05}$ | $1.11^{0.01}_{0.09}$ | $0.86^{0.12}_{0.16}$ | $1.53^{0.31}_{0.1}$ | $0.43^{0.12}_{0.25}$ |
| 2M13472354 | $0.33^{0.04}_{0.09}$ | $<1.46$ | ... | $<0.71$ | $0.59^{0.11}_{0.3}$ | $0.86^{0.03}_{0.09}$ | $0.87^{0.12}_{0.24}$ | $1.65^{0.26}_{0.1}$ | $0.56^{0.13}_{0.1}$ |
| 2M13535604 | $-0.06^{0.02}_{0.68}$ | $<0.77$ | $<1.56$ | $<0.28$ | $0.91^{0.1}_{0.24}$ | $1.75^{0.01}_{0.09}$ | $1.39^{0.13}_{0.09}$ | $2.1^{0.25}_{0.1}$ | $0.5^{0.13}_{0.09}$ |
| 2M16092248 | $-0.05^{0.11}_{0.13}$ | $<0.83$ | $1.17^{0.06}_{0.29}$ | $<0.35$ | $0.68^{0.1}_{0.01}$ | $1.29^{0}_{0.1}$ | $1.02^{0.13}_{0.02}$ | $1.81^{0.21}_{0.09}$ | $0.12^{0.19}_{0.2}$ |
| 2M16482145 | $0.11^{0.14}_{0.35}$ | $<0.41$ | $0.89^{0.05}_{0.11}$ | $<0.23$ | $0.67^{0.15}_{0.04}$ | $1.19^{0.07}_{0.09}$ | $0.79^{0.14}_{0.16}$ | $1.57^{0.17}_{0.1}$ | $0.41^{0.21}_{0.46}$ |
| 2M17033361 | $0.06^{0.08}_{0.12}$ | $0.41^{0.2}_{0.09}$ | $<0.71$ | $<0.14$ | $0.68^{0.12}_{0.01}$ | $1.22^{0.06}_{0.1}$ | $0.73^{0.14}_{0.02}$ | $1.8^{0.31}_{0.09}$ | $0.03^{0.2}_{0.1}$ |
| 2M17183459 | $0.07^{0.19}_{0.02}$ | $<1.17$ | $<1.21$ | $<0.37$ | $0.67^{0.08}_{0.04}$ | $1.48^{0.01}_{0.1}$ | $1^{0.14}_{0.01}$ | $1.37^{0.03}_{0.09}$ | $0.34^{0.01}_{0.03}$ |
| 2M17214096 | $-0.11^{0.14}_{0.04}$ | $<0.72$ | $1^{0.06}_{0.04}$ | $<0.48$ | $0.64^{0.14}_{0.01}$ | $1.32^{0.04}_{0.09}$ | $1.02^{0.12}_{0.04}$ | $2.04^{0.02}_{0.1}$ | $0.44^{0.09}_{0.11}$ |
| 2M17502038 | $-0.12^{0.09}_{0.09}$ | $<0.61$ | $1.16^{0.07}_{0.1}$ | $<0.16$ | $0.7^{0.1}_{0.1}$ | $1.42^{0.17}_{0.1}$ | $0.75^{0.11}_{0}$ | $1.69^{0.1}_{0.1}$ | $0.27^{0.2}_{0.1}$ |
| 2M18120031 | $0.01^{0.12}_{0.1}$ | $0.97^{0.07}_{0.47}$ | $0.96^{0.04}_{0.1}$ | $<0.12$ | $0.57^{0.12}_{0.05}$ | $1.22^{0.03}_{0.1}$ | $0.76^{0.12}_{0.1}$ | $1.65^{0.08}_{0.1}$ | $0.35^{0.16}_{0.17}$ |
| 2M19105369 | $0.04^{0.2}_{0.84}$ | $1.42^{0.19}_{0.64}$ | $<0.97$ | $<0.27$ | $0.61^{0.11}_{0.09}$ | $1.1^{0.01}_{0.09}$ | $0.87^{0.11}_{0.03}$ | $1.69^{0.23}_{0.1}$ | $0.49^{0.2}_{0.05}$ |
| 2M19214936 | $0.05^{0.05}_{0.07}$ | $0.6^{0.12}_{0.05}$ | $0.91^{0.05}_{0.05}$ | $0.12^{0.13}_{0.1}$ | $0.53^{0.08}_{0.09}$ | $0.93^{0.01}_{0.09}$ | $0.71^{0.12}_{0.09}$ | $1.44^{0.24}_{0.1}$ | $0.21^{0.07}_{0.19}$ |
| 2M19281906 | $-0.11^{0.11}_{0.08}$ | $<0.7$ | $0.73^{0.05}_{0.13}$ | $<0.56$ | $0.47^{0.13}_{0.02}$ | $0.89^{0.05}_{0.09}$ | $0.72^{0.11}_{0.1}$ | $1.28^{0.07}_{0}$ | $0.24^{0.07}_{0}$ |

| star | [K/Fe] | [Ca/Fe] | [Ti/Fe] | [V/Fe] | [Cr/Fe] | [Mn/Fe] | [Co/Fe] | [Ni/Fe] | [Ce/Fe] |
|---|---|---|---|---|---|---|---|---|---|
| 2M00044180 | $0.08^{0.03}_{0.01}$ | $0.27^{0.02}_{0.03}$ | $0.24^{0.03}_{0.01}$ | $<0.72$ | $<0.22$ | $-0.42^{0.05}_{0.01}$ | $<0.15$ | $0.11^{0.07}_{0.06}$ | $<0.02$ |
| 2M12071315 | $0.35^{0.02}_{0.27}$ | $0.19^{0.08}_{0.01}$ | $0.25^{0.01}_{0.14}$ | $<1.78$ | $<-0.06$ | $-0.2^{0.11}_{0.1}$ | $<0.43$ | $0.06^{0.13}_{0}$ | $0.67^{0.09}_{0.33}$ |
| 2M13314691 | $0.06^{0.06}_{0.09}$ | $0.31^{0.02}_{0.09}$ | $0.4^{0.1}_{0.26}$ | $0.23^{0.04}_{0.1}$ | $0.04^{0.05}_{0.01}$ | $-0.33^{0.04}_{0}$ | $0.09^{0.11}_{0.02}$ | $0.09^{0.05}_{0.03}$ | $0.79^{0.08}_{0.17}$ |
| 2M13472354 | $<0.36$ | $0.12^{0.05}_{0.08}$ | $<0.58$ | ... | $<0.56$ | $<0.19$ | $<1.47$ | $0.18^{0.1}_{0.1}$ | $<0.77$ |
| 2M13535604 | $0.16^{0.03}_{0.19}$ | $0.28^{0.1}_{0.06}$ | $0.16^{0.07}_{0.1}$ | $<0.89$ | $<0.25$ | $-0.06^{0.24}_{0.19}$ | $<0.51$ | $0.28^{0.12}_{0.01}$ | $0.85^{0.1}_{0.09}$ |
| 2M16092248 | $0.09^{0}_{0.04}$ | $0.15^{0.07}_{0.01}$ | $0.32^{0.2}_{0.05}$ | $<0.17$ | $<-0.07$ | $<-0.45$ | $0.21^{0.08}_{0.04}$ | $0.2^{0.05}_{0}$ | $0.78^{0.11}_{0.09}$ |
| 2M16482145 | $0.12^{0.11}_{0.01}$ | $0.25^{0.05}_{0.07}$ | $0.43^{0.16}_{0.09}$ | $<0.45$ | $<-0.03$ | $-0.11^{0.07}_{0.01}$ | $0.04^{0.12}_{0}$ | $0.02^{0.05}_{0.1}$ | $0.68^{0.09}_{0.58}$ |
| 2M17033361 | $0.17^{0.01}_{0.11}$ | $0.38^{0.13}_{0.01}$ | $0.7^{0.21}_{0.08}$ | $<0.71$ | $<-0.16$ | $0.00^{0.25}_{0.19}$ | $0.28^{0.15}_{0.04}$ | $0.02^{0.24}_{0}$ | $1.04^{0.14}_{0.04}$ |
| 2M17183459 | $0.06^{0.05}_{0.06}$ | $0.27^{0.1}_{0}$ | $0.77^{0.15}_{0.07}$ | $<1.23$ | $<0.27$ | $-0.18^{0.13}_{0.02}$ | $<0.48$ | $0.07^{0.12}_{0.1}$ | $0.68^{0.23}_{0.3}$ |
| 2M17214096 | $0.16^{0.03}_{0.07}$ | $0.04^{0.11}_{0.08}$ | $0.32^{0.1}_{0.1}$ | $<0.63$ | $<0.14$ | $-0.26^{0.05}_{0.03}$ | $0.19^{0.02}_{0.01}$ | $0.1^{0.05}_{0.04}$ | $0.79^{0.1}_{0.09}$ |
| 2M17502038 | $0.38^{0.06}_{0.08}$ | $0.48^{0.03}_{0.08}$ | $0.49^{0.15}_{0.06}$ | $0.43^{0.06}_{0.03}$ | $0.12^{0.05}_{0.24}$ | $-0.28^{0.07}_{0.06}$ | $0.06^{0.04}_{0.02}$ | $-0.19^{0.19}_{0.27}$ | $0.57^{0.15}_{0.1}$ |
| 2M18120031 | $0.21^{0.02}_{0.02}$ | $0.34^{0.02}_{0.06}$ | $0.45^{0.14}_{0.25}$ | $0.28^{0.07}_{0.1}$ | $<0.14$ | $-0.21^{0.3}_{0.1}$ | $0.07^{0.06}_{0.1}$ | $-0.03^{0.11}_{0.12}$ | $0.74^{0.17}_{0.1}$ |
| 2M19105369 | $0.26^{0.05}_{0.07}$ | $0.23^{0.15}_{0.05}$ | $0.55^{0.2}_{0.02}$ | $<2.06$ | $<0.19$ | $-0.1^{0.12}_{0.01}$ | $<0.5$ | $0.1^{0.13}_{0.1}$ | $1.11^{0.05}_{0.09}$ |
| 2M19214936 | $0.11^{0.01}_{0.12}$ | $0.31^{0.01}_{0.13}$ | $0.36^{0.19}_{0.14}$ | $0.17^{0.07}_{0.12}$ | $-0.09^{0.08}_{0.05}$ | $-0.23^{0.2}_{0.27}$ | $0.16^{0.02}_{0.04}$ | $-0.01^{0.06}_{0.04}$ | $0.7^{0.13}_{0.15}$ |
| 2M19281906 | $0.13^{0.04}_{0.07}$ | $0.19^{0.02}_{0.01}$ | $0.38^{0.2}_{0.1}$ | $<0.63$ | $-0.03^{0.27}_{0.08}$ | $-0.05^{0.29}_{0.1}$ | $0.13^{0.06}_{0.03}$ | $0.15^{0.04}_{0.02}$ | $0.75^{0.13}_{0.22}$ |

All abundances are in logarithmic scale, relative to their metallicity and to the Solar values[1] such as displayed in Fig.1, 4 and 6 . The upper indices show the random errors while the bottom indices show the systematic errors.

Supplementary Table 4: Abundances of the P-normal stars

| star | [C/Fe] | [N/Fe] | [O/Fe] | [Na/Fe] | [Mg/Fe] | [Al/Fe] | [Si/Fe] | [P/Fe] | [S/Fe] |
|---|---|---|---|---|---|---|---|---|---|
| 2M00022364 | $0.24_{0.13}^{0.19}$ | $0.38_{0.1}^{0.07}$ | $<0.56$ | $<0.29$ | $0.43_{0.11}^{0.08}$ | $0.45_{0.17}^{0.12}$ | $0.37_{0.05}^{0.09}$ | $<0.51$ | $0.26_{0.14}^{0.17}$ |
| 2M01262518 | $0.29_{0.02}^{0.18}$ | $0.2_{0.1}^{0.03}$ | $0.52_{0.1}^{0.17}$ | $<0.47$ | $0.4_{0.08}^{0.07}$ | $0.45_{0.14}^{0.09}$ | $0.34_{0.02}^{0.09}$ | $<0.32$ | $0.27_{0.05}^{0.17}$ |
| 2M01271266 | $0.09_{0.15}^{0.06}$ | $0.17_{0.06}^{0.03}$ | $0.61_{0.09}^{0.11}$ | $0.02_{0.1}^{0.18}$ | $0.4_{0.03}^{0.11}$ | $0.36_{0.02}^{0.02}$ | $0.3_{0.03}^{0.07}$ | $0.41_{0.09}^{0.28}$ | $0.34_{0.1}^{0.17}$ |
| 2M02150319 | $0.32_{0.23}^{0.03}$ | $0.23_{0.48}^{0.22}$ | $0.61_{0.04}^{0.05}$ | $0.46_{0.04}^{0.19}$ | $0.41_{0.04}^{0.09}$ | $0.37_{0.00}^{0.00}$ | $0.25_{0.04}^{0.07}$ | $<-0.25$ | $0.18_{0.04}^{0.11}$ |
| 2M07404920 | $0.13_{0.12}^{0.09}$ | $0.19_{0.06}^{0.08}$ | $0.54_{0.09}^{0.04}$ | $0.09_{0.04}^{0.14}$ | $0.34_{0.04}^{0.13}$ | $0.44_{0.00}^{0.03}$ | $0.3_{0.03}^{0.07}$ | $<0.67$ | $0.27_{0.16}^{0.08}$ |
| 2M07461742 | $0.22_{0.06}^{0.13}$ | $0.24_{0.15}^{0.05}$ | $0.6_{0.11}^{0.03}$ | $<0.38$ | $0.33_{0.06}^{0.05}$ | $0.36_{0.00}^{0.05}$ | $0.24_{0.03}^{0.06}$ | $<0.66$ | $0.14_{0.03}^{0.1}$ |
| 2M11202609 | $0.29_{0.15}^{0.03}$ | $0.46_{0.01}^{0.04}$ | $0.5_{0.04}^{0.05}$ | $0.53_{0.06}^{0.12}$ | $0.4_{0.04}^{0.13}$ | $-0.02_{0.00}^{0.03}$ | $0.25_{0.08}^{0.1}$ | $<0.59$ | $0.45_{0.06}^{0.2}$ |
| 2M13240275 | $0.15_{0.21}^{0.24}$ | $0.21_{0.00}^{0.15}$ | $<0.46$ | $<0.39$ | $0.41_{0.08}^{0.1}$ | $0.38_{0.11}^{0.11}$ | $0.33_{0.06}^{0.1}$ | $0.33_{0.09}^{0.33}$ | $0.32_{0.06}^{0.25}$ |
| 2M15563661 | $0.04_{0.00}^{0.08}$ | $0.29_{0.05}^{0.03}$ | $0.56_{0.03}^{0.13}$ | $0.28_{0.1}^{0.25}$ | $0.4_{0.09}^{0.08}$ | $0.45_{0.05}^{0.05}$ | $0.29_{0.00}^{0.07}$ | $<0.11$ | $0.18_{0.07}^{0.02}$ |
| 2M16164586 | $0.00_{0.09}^{0.11}$ | $0.18_{0.06}^{0.04}$ | $0.59_{0.04}^{0.11}$ | $<0.29$ | $0.42_{0.00}^{0.08}$ | $0.38_{0.02}^{0.00}$ | $0.32_{0.01}^{0.06}$ | ... | $0.26_{0.04}^{0.06}$ |
| 2M16471103 | $-0.39_{0.05}^{0.12}$ | $1.48_{0.1}^{0.03}$ | $0.53_{0.55}^{0.1}$ | $0.39_{0.09}^{0.12}$ | $0.35_{0.2}^{0.08}$ | $0.44_{0.03}^{0.26}$ | $0.26_{0.06}^{0.09}$ | ... | $0.11_{0.08}^{0.07}$ |
| 2M17142525 | $-0.2_{0.00}^{0.14}$ | $0.2_{0.11}^{0.04}$ | $0.55_{0.1}^{0.06}$ | $0.22_{0.1}^{0.17}$ | $0.31_{0.00}^{0.1}$ | $0.01_{0.02}^{0.09}$ | $0.25_{0.04}^{0.07}$ | $<0.58$ | $0.25_{0.15}^{0.15}$ |
| 2M19311218 | $0.15_{0.00}^{0.06}$ | $0.38_{0.02}^{0.07}$ | $0.6_{0.03}^{0.08}$ | $<-0.02$ | $0.42_{0.00}^{0.1}$ | $0.55_{0.00}^{0.02}$ | $0.39_{0.00}^{0.09}$ | $0.47_{0.1}^{0.2}$ | $0.33_{0.02}^{0.11}$ |
| 2M19320870 | $0.16_{0.06}^{0.06}$ | $<0.18$ | $0.62_{0.14}^{0.08}$ | $<0.1$ | $0.39_{0.02}^{0.11}$ | $0.5_{0.1}^{0.04}$ | $0.29_{0.04}^{0.08}$ | $0.78_{0.09}^{0.2}$ | $0.33_{0.15}^{0.1}$ |
| 2M21330531 | $0.09_{0.13}^{0.06}$ | $<0.25$ | $0.65_{0.01}^{0.06}$ | $<0.16$ | $0.4_{0.08}^{0.1}$ | $0.54_{0.1}^{0.03}$ | $0.44_{0.01}^{0.06}$ | $<0.84$ | $0.27_{0.11}^{0.07}$ |

| star | [K/Fe] | [Ca/Fe] | [Ti/Fe] | [V/Fe] | [Cr/Fe] | [Mn/Fe] | [Co/Fe] | [Ni/Fe] | [Ce/Fe] |
|---|---|---|---|---|---|---|---|---|---|
| 2M00022364 | $0.19_{0.04}^{0.00}$ | $0.15_{0.03}^{0.04}$ | $0.14_{0.74}^{0.07}$ | ... | $-0.2_{0.1}^{0.1}$ | $-0.41_{0.06}^{0.05}$ | $<-0.1$ | $0.03_{0.11}^{0.12}$ | $0.25_{0.1}^{0.32}$ |
| 2M01262518 | $0.06_{0.16}^{0.07}$ | $0.06_{0.00}^{0.01}$ | $0.09_{0.86}^{0.2}$ | $<1.8$ | $-0.28_{0.1}^{0.2}$ | $-0.28_{0.1}^{0.25}$ | $<0.12$ | $0.05_{0.11}^{0.04}$ | $<0.11$ |
| 2M01271266 | $0.15_{0.28}^{0.05}$ | $0.3_{0.00}^{0.04}$ | $0.38_{0.06}^{0.2}$ | $<-0.88$ | $-0.08_{0.1}^{0.00}$ | $-0.37_{0.12}^{0.19}$ | $0.13_{0.1}^{0.18}$ | $0.05_{0.04}^{0.04}$ | $0.24_{0.1}^{0.1}$ |
| 2M02150319 | $0.14_{0.06}^{0.06}$ | $0.17_{0.06}^{0.01}$ | $0.11_{0.05}^{0.09}$ | $<0.34$ | $0.1_{0.04}^{0.00}$ | $-0.32_{0.1}^{0.11}$ | $0.14_{0.1}^{0.18}$ | $0.18_{0.01}^{0.08}$ | $<-0.2$ |
| 2M07404920 | $0.18_{0.03}^{0.03}$ | $0.16_{0.05}^{0.02}$ | $0.29_{0.13}^{0.16}$ | $<0.36$ | $-0.13_{0.01}^{0.06}$ | $-0.34_{0.14}^{0.03}$ | $0.07_{0.01}^{0.12}$ | $0.07_{0.06}^{0.00}$ | $<-0.18$ |
| 2M07461742 | $0.18_{0.38}^{0.07}$ | $0.22_{0.02}^{0.05}$ | $0.12_{0.13}^{0.14}$ | $<0.55$ | $0.03_{0.36}^{0.06}$ | $-0.18_{0.1}^{0.05}$ | $0.13_{0.1}^{0.1}$ | $0.13_{0.06}^{0.05}$ | $0.1_{0.1}^{0.14}$ |
| 2M11202609 | $0.27_{0.16}^{0.00}$ | $0.26_{0.05}^{0.02}$ | $0.63_{0.44}^{0.23}$ | $<-0.01$ | $0.00_{0.1}^{0.07}$ | $0.2_{0.46}^{0.39}$ | $-0.13_{0.18}^{0.08}$ | $0.05_{0.02}^{0.03}$ | $0.94_{0.09}^{0.15}$ |
| 2M13240275 | $0.11_{0.26}^{0.00}$ | $0.12_{0.05}^{0.01}$ | $0.2_{0.76}^{0.15}$ | $<0.67$ | $-0.07_{0.1}^{0.11}$ | $-0.37_{0.07}^{0.07}$ | $<-0.1$ | $0.08_{0.1}^{0.07}$ | $0.02_{0.1}^{0.13}$ |
| 2M15563661 | $0.09_{0.08}^{0.04}$ | $0.22_{0.06}^{0.03}$ | $0.37_{0.09}^{0.28}$ | $<-0.08$ | $-0.08_{0.28}^{0.24}$ | $-0.35_{0.08}^{0.00}$ | $0.11_{0.05}^{0.12}$ | $0.03_{0.04}^{0.05}$ | $0.13_{0.13}^{0.06}$ |
| 2M16164586 | $0.14_{0.23}^{0.05}$ | $0.28_{0.00}^{0.00}$ | $0.15_{0.07}^{0.07}$ | $0.24_{0.1}^{0.11}$ | $0.14_{0.1}^{0.04}$ | $-0.34_{0.02}^{0.07}$ | $-0.28_{0.1}^{0.15}$ | $0.05_{0.04}^{0.03}$ | $0.41_{0.03}^{0.03}$ |
| 2M16471103 | $0.16_{0.3}^{0.01}$ | $0.23_{0.12}^{0.03}$ | $0.56_{0.67}^{0.19}$ | $<0.16$ | $0.14_{0.1}^{0.13}$ | $-0.32_{0.34}^{0.01}$ | $-0.27_{0.1}^{0.09}$ | $0.04_{0.08}^{0.04}$ | $0.37_{0.09}^{0.02}$ |
| 2M17142525 | $0.11_{0.05}^{0.02}$ | $0.23_{0.01}^{0.01}$ | $0.11_{0.42}^{0.06}$ | $<3.11$ | $0.01_{0.03}^{0.08}$ | $-0.2_{0.06}^{0.28}$ | $-0.16_{0.1}^{0.13}$ | $-0.04_{0.06}^{0.07}$ | $0.17_{0.07}^{0.08}$ |
| 2M19311218 | $0.12_{0.00}^{0.03}$ | $0.14_{0.00}^{0.03}$ | $0.47_{0.04}^{0.26}$ | $<0.58$ | $0.00_{0.00}^{0.16}$ | $-0.35_{0.00}^{0.12}$ | $<-0.05$ | $0.09_{0.00}^{0.07}$ | $<-0.21$ |
| 2M19320870 | $0.09_{0.11}^{0.02}$ | $0.26_{0.03}^{0.03}$ | $0.37_{0.4}^{0.31}$ | $<0.62$ | $0.14_{0.14}^{0.09}$ | $-0.44_{0.2}^{0.06}$ | $0.17_{0.01}^{0.01}$ | $0.04_{0.04}^{0.09}$ | $<-0.05$ |
| 2M21330531 | $0.05_{0.02}^{0.01}$ | $0.29_{0.04}^{0.06}$ | $0.18_{0.09}^{0.02}$ | $<0.47$ | $-0.06_{0.1}^{0.18}$ | $-0.27_{0.2}^{0.26}$ | $0.11_{0.03}^{0.06}$ | $0.07_{0.01}^{0.07}$ | $0.39_{0.29}^{0.07}$ |

All abundances are in logarithmic scale, relative to their metallicity and to the Solar values[1] such as displayed in Fig.1, 4 and 6. The upper indices show the random errors while the bottom indices show the systematic errors.

# References

[1] Asplund, M., Grevesse, N., Sauval, A. J., Scott, P. The Chemical Composition of the Sun. Ann. Rev. Astron. Astrophys. **47**, 481-522 (2009)